\documentclass[11pt,english,superscriptaddress,aps,preprint]{revtex4}

\usepackage{amsmath}
\usepackage{amssymb}
\usepackage{graphicx}
\usepackage{xcolor}

\makeatletter

\usepackage{slashed}

\makeatother

\usepackage{babel}

\begin{document}

\title {Field sources in a planar anisotropic CPT-odd gauge model}

\author{L.H.C. Borges}
\email{luizhenrique.borges@ufla.br}

\affiliation{Departamento de Física, Universidade Federal de Lavras, Caixa Postal 3037, 37200-900 Lavras-MG, Brazil}

\author{A.F. Ferrari}
\email{alysson.ferrari@ufabc.edu.br}

\author{P.H.O. da Silva}
\email{silva.pedro@aluno.ufabc.edu.br}

\affiliation{Universidade Federal do ABC, Centro de Ci\^encias Naturais e Humanas, Rua Santa Ad\'elia, 166, 09210-170, Santo Andr\'e, SP, Brasil}

\author{F. A. Barone}
\email{fbarone@unifei.edu.br}

\affiliation{IFQ - Universidade Federal de Itajubá, Av. BPS 1303, Pinheirinho, Caixa Postal 50, 37500-903, Itajubá, MG, Brazil}

\begin{abstract}
In the present paper we study some new classical Lorentz violating effects in planar electrodynamics due to the presence of stationary point-like field sources. Starting from the Carroll-Field-Jackiw model defined in $(3+1)$ dimensions, which belongs to the electromagnetic CPT-odd sector of the Standard Model Extension (SME), we apply the dimensional reduction procedure obtaining a $(2+1)$-dimensional model that encompasses an electromagnetic sector composed of the Maxwell-Chern-Simons electrodynamics, a pure scalar sector described by a massless Klein-Gordon field, and a mixed sector where the background vector mediates contributions involving both the scalar and the gauge fields. For all the sectors of this planar theory, we explore some physical phenomena that arise from the interactions between external sources. Specifically, we obtain perturbative effects up to second order in the background vector related to the presence of both electric and scalar planar charges and Dirac points.
\end{abstract}

\maketitle

\section{INTRODUCTION}
\label{INTRO}

Gauge models can be powerful tools to establish effective theories for describing condensed matter systems. In this scenario, we highlight the planar models \cite{MPLA2021Campos2150099,PRD2004Dalmazi065021,JPA2006Dalmazi39,PLA2003Paschoal313,PLA2003Colatto314,PRD2001Irvine045015,PRD1999Carrington125018,PRD1995Carrington1903} and their application in the study of surface, interface, and thin material physics \cite{PRB2016Sinner2016,PRD2020Magalhaes101,Chaos2000Escalona337,PRD1997Escalona6327,PRD1995Carrington4451,PRD1995Burgess5052,PRD1994Carrington2830,PLB1990Doreya250}. In particular, the Abelian Maxwell-Chern-Simons-type models, which represent topologically massive gauge theories \cite{MCS1}, are commonly employed in planar condensed matter physics (see, for instance, \cite{MCS2} and references therein).

The anisotropic properties which can be found in material surfaces and interfaces are one of the most frequently addressed subjects in the literature, regarding planar systems, mainly the ones related to the phonon propagation \cite{PRAP2020Neogi024004,RR2015Wang3944,SR2021Hoang21202,PRB2021Hongo134403,AOM2022Gong2200023,2DM2023Muhammad025001,AES2023Coco100135,PC1990Tewari647,FRP2023Cai43303,N2016Qin11306}. Effective field models employed to describe this type of phenomea must encompass anisotropy and exhibit breaking of rotational symmetry, among others. We can find a vast literature of such kind of models in the issue of Lorentz symmetry breaking and we would like to draw special attention for the work of reference \cite{pSMEodd}, which is obtained by dimensional reduction of the so called Carroll-Field-Jackiw Electrodynamics \cite{CFJ}.

Since the proposal of the $(3+1)$-dimensional Standard Model Extension (SME) \cite{SME1,SME2}, Lorentz violating field theories have been extensively explored in the literature, mainly in the gauge sector, which is composed of both the CPT-odd and CPT-even parts. In the Abelian (Maxwell) sector, the CPT-odd part is described by the Carroll-Field-Jackiw term $\epsilon^{{{\mu}}{{\nu}}{{\kappa}}{{\lambda}}}\left(k_{AF}\right)_{{{\mu}}}A_{{{\nu}}}F_{{{\kappa}}{{\lambda}}}$ \cite{CFJ}, while the CPT-even one corresponds to the  $\left(k_{F}\right)_{{{\kappa}}{{\lambda}}{{\mu}}{{\nu}}}F^{{{\kappa}}{{\lambda}}}F^{{{\mu}}{{\nu}}}$ term \cite{SME1}. As some interesting developments, we can mention the search for Lorentz violation effects in electromagnetic wave propagation \cite{wave1,wave2,wave3,wave4,wave5,wave6,wave7,wave8}, classical electrodynamics \cite{ce1,ce2,ce3,ce4,ce5}, QED \cite{qed1,qed2,qed3,qed4,qed5}, models subject to boundary conditions \cite{bc1,bc2,bc3,bc4}, atomic physics \cite{ap1,ap2,ap3,ap4,ap5}, thermal quantum field theory \cite{t1,t2,t3,t4}, among others. 

Lower-dimensional physics can be an interesting place not only for the search of Lorentz violating effects in condensed matter systems, but also to employ Lorentz violating field models to establish effective descriptions of material media. One of the ways to look for such effects in electrodynamics is through the planar version of the SME, which is obtained from the $(3+1)$-dimensional SME by applying the dimensional reduction procedure. The planar version of the Abelian gauge sector of the SME was obtained for both the CPT-odd \cite{pSMEodd} and CPT-even \cite{pSMEeven} parts, and these results were recently extended to include nonminimal (higher derivative) terms \cite{pnmSME}. In the framework of the planar SME, we mention, for instance, investigations which concern stability, causality and unitarity properties \cite{pSMEodd,sucpodd,sucpoeven}, classical solutions \cite{pSMEeven,clodd,cloddp} and Aharonov-Bohm scattering \cite{AB}. We also highlight investigations which concern the planar electrodynamics modified by Lorentz violating higher-derivative terms \cite{MMFpalanar} .

Regarding the interactions between external sources in a Lorentz violating context, we highlight investigations in the electromagnetic sector of the $(3+1)$-dimensional SME \cite{ce1,ce2,fontesMPLA}. However, studies of this type have not yet been performed in the planar electrodynamics. This topic is an interesting subject not only because it brings the possibility of searching for Lorentz violation signals in lower-dimensional condensed matter systems, but also because these models can be employed to the study physical properties of planar materials. Besides, anisotropies are usual features of many condensed matter systems and models which exhibit Lorentz-symmetry breaking can be a convenient tool in this scenario.

Inspired by this, in this paper we search for physical phenomena due to the presence of stationary point-like field sources in CPT-odd planar electrodynamics. 

From the Carroll-Field-Jackiw model defined in $(3+1)$ dimensions, we apply the dimensional reduction procedure described in Ref. \cite{pSMEodd}, resulting in a $(2+1)$-dimensional theory that includes the Maxwell-Chern-Simons electrodynamics, a pure scalar sector, described by a massless Klein-Gordon field, and a mixed sector where the background vector mediates contributions involving both the scalar and the gauge fields. For all the sectors of this planar theory, we investigate some physical phenomena that emerge from the interactions between external sources. Specifically, we obtain perturbative results up to second order in the background vector related to the presence of both electric and scalar planar charges and Dirac points. In general, we obtain physical phenomena such as spontaneous torques, interaction forces and electromagnetic fields. In the absence of Lorentz violation, in the electromagnetic sector, we recover the known results in the usual Maxwell-Chern-Simons electrodynamics \cite{sourcesMCS}.

The paper is organized as follows: In section \ref{VENERGY} we compute the contributions, due to the sources, to the vacuum energy of the system. In section \ref{charges}, we obtain effects due to the presence of both electric and scalar planar charges. In section \ref{DIRAC}, we explore some physical phenomena involving Dirac points and, finally, the electromagnetic field configurations generated for both the electric planar charge and the Dirac point are studied in section \ref{EM}. Section \ref{conclusions} is devoted to our final remarks and conclusions. 

From here on, we will adopt the following conventions: for the $(3 + 1)$-dimensional spacetime, we will use hatted greek indices, and the Minkowski metric is $\eta^{{\hat{\mu}}\hat{\nu}}=\left(+, -, -, -\right)$ with ${\hat{\mu}}=0,1,2,3$. The Levi-Civita tensor is denoted by $\epsilon^{{\hat{\mu}}{\hat{\nu}}{\hat{\kappa}}{\hat{\lambda}}}$ where $\epsilon^{0123}=1$.  The analogue quantities for the $(2+1)$-dimensional spacetime are $\eta^{\mu\nu}=\left(+, -, -\right)$, $\mu=0,1,2$ and we have $\epsilon^{\mu\nu\alpha}$ with $\epsilon^{012}=1$.

\section{VACUUM ENERGY IN THE PRESENCE OF FIELD SOURCES }
\label{VENERGY}

We start this section by writing the Lagrangian density which describes the gauge CPT-odd sector of the SME in $3+1$ dimensions \cite{SME1},
\begin{eqnarray}
\label{lagCPTodd}
{\cal{L}}_{(3+1)}=-\frac{1}{4}F_{{\hat{\mu}}{\hat{\nu}}}F^{{\hat{\mu}}{\hat{\nu}}}+\frac{1}{2}\epsilon^{{\hat{\mu}}{\hat{\nu}}{\hat{\kappa}}{\hat{\lambda}}}\left(k_{AF}\right)_{{\hat{\mu}}}A_{{\hat{\nu}}}F_{{\hat{\kappa}}{\hat{\lambda}}}+J^{{\hat{\mu}}}A_{{\hat{\mu}}} \ ,
\end{eqnarray}
where $A^{{\hat{\mu}}}$ is the gauge field, $F^{{\hat{\mu}}{\hat{\nu}}}=\partial^{{\hat{\mu}}}A^{{\hat{\nu}}}-\partial^{{\hat{\nu}}}A^{{\hat{\mu}}}$ stands for the field strength, $J^{{\hat{\mu}}}$ is an electromagnetic field source and $\left(k_{AF}\right)^{{\hat{\mu}}}$ is a constant background vector with mass dimension, responsible for the Lorentz violation.

In order to obtain the planar version of the model (\ref{lagCPTodd}), we adopt the dimensional reduction  prescription described in Ref. \cite{pSMEodd}, which implies in the following replacements,
\begin{eqnarray}
\label{RDime}
A^{{\hat{\mu}}}\rightarrow\left(A^{\mu};\phi\right) \ , \  \left(k_{AF}\right)^{{\hat{\mu}}}\rightarrow\frac{1}{2}\left(v^{\mu};m\right) \ , \ J^{{\hat{\mu}}}\rightarrow\left(J^{\mu};J\right) \ ,
\end{eqnarray}
where $A^{3}=\phi \ , \ \left(k_{AF}\right)^{3}=\frac{m}{2} \ , \ J^{3}=J$. Thereby, the corresponding $(2+1)$-dimensional model reads,  
\begin{eqnarray}
\label{lagCPToddRd}
{\cal{L}}_{(2+1)}&=&-\frac{1}{4}F_{\mu\nu}F^{\mu\nu}+\frac{m}{2}\epsilon_{\mu\nu\kappa}A^{\mu}\partial^{\nu}A^{\kappa}-\frac{1}{2\alpha}\left(\partial_{\mu}A^{\mu}\right)^{2}+J^{\mu}A_{\mu}+\frac{1}{2}\partial_{\mu}\phi\partial^{\mu}\phi - J\phi\nonumber\\
&
&-\frac{1}{2}\phi\epsilon_{\mu\nu\kappa}v^{\mu}\partial^{\nu}A^{\kappa}+\frac{1}{2}\epsilon_{\mu\nu\kappa}v^{\mu}\left(\partial^{\nu}\phi\right)A^{\kappa} \ . 
\end{eqnarray}
As a result of this dimensional reduction we have an electromagnetic sector governed by the Maxwell-Chern-Simons electrodynamics, with $m$ standing for a mass parameter, a pure scalar sector, which is described by the Klein-Gordon massless field $\phi$, and a mixing sector, where the background vector $v^{\mu}$ mediates contributions involving both the fields $A^{\mu}$ and $\phi$. The third term in (\ref{lagCPToddRd}) fixes the gauge, and $J^{\mu}$ and $J$ are the external sources for the gauge and scalar fields, respectively.

In this work we shall focus on the lagrangian (\ref{lagCPToddRd}) disregarding its relation to the higher dimensional SME.

The functional generator for the Lagrangian (\ref{lagCPToddRd}) can be written in the matrix form,
\begin{eqnarray}
\label{gfun}
{\cal{Z}}_{(2+1)}&=&\int{\cal{D}}A {\cal{D}}\phi\exp\Biggl\{\int d^{3}x\Biggl[\frac{i}{2}\bordermatrix{&  \cr & A^{\mu}\left(x\right) \  & \phi\left(x\right)  \   \cr }\bordermatrix{&  \cr & {\cal{O}}_{\mu\nu}\left(x\right) \ \ & {\cal{O}}_{\mu}\left(x\right)  \   \cr & -{\cal{O}}_{\nu}\left(x\right) \ \ & -\Box\left(x\right) \ \cr }\bordermatrix{&  \cr & A^{\nu}\left(x\right) \ \ \cr & \phi\left(x\right)  \   \cr }\nonumber\\
&
&+i\bordermatrix{&  \cr & A_{\mu}\left(x\right) \  & \phi\left(x\right)  \   \cr }\bordermatrix{&  \cr & J^{\mu}\left(x\right) \ \ \cr & -J \left(x\right) \   \cr }
\Biggr]\Biggr\} \ ,
\end{eqnarray}
where we defined the differential operators,
\begin{eqnarray}
\label{OperaD}
{\cal{O}}_{\mu\nu}\left(x\right)&=&\Box\eta_{\mu\nu}-\left(1-\frac{1}{\alpha}\right)\partial_{\mu}\partial_{\nu}+m\epsilon_{\mu\kappa\nu}\partial^{\kappa} \ , \ {\cal{O}}_{\mu}\left(x\right)=\epsilon_{\nu\alpha\mu}v^{\nu}\partial^{\alpha} \ , \nonumber\\
&
&\ {\cal{O}}_{\nu}\left(x\right)=\epsilon_{\mu\alpha\nu}v^{\mu}\partial^{\alpha} \ , \ \Box\left(x\right)= \partial_{\beta}\partial^{\beta} \ .
\end{eqnarray}

Carrying out the Gaussian integral (\ref{gfun}), we arrive at
\begin{eqnarray}
\label{gfunGau}
{\cal{Z}}_{(2+1)}=\exp\Bigg[-\frac{i}{2}\int\int d^{3}x d^{3}y\cr\cr
\bordermatrix{&  \cr & J^{\mu}\left(x\right) \  & -J\left(x\right)  \   \cr }\bordermatrix{&  \cr & D_{\mu\nu}\left(x,y\right) \ \ & G_{\mu}\left(x,y\right)  \   \cr & \Delta_{\nu}\left(x,y\right) \ \ & D\left(x,y\right) \ \cr }
\bordermatrix{&  \cr & J^{\nu}\left(y\right) \ \ \cr & -J \left(y\right) \   \cr }\Bigg] , \nonumber\\ 
\end{eqnarray}
where we defined de propagator as a Fourier integral, 
\begin{eqnarray}
\label{prop1}
\bordermatrix{&  \cr & D_{\mu\nu}\left(x,y\right) \ \ & G_{\mu}\left(x,y\right)  \   \cr & \Delta_{\nu}\left(x,y\right) \ \ & D\left(x,y\right) \ \cr }=\int\frac{d^{3}p}{\left(2\pi\right)^{3}}\bordermatrix{&  \cr & {\tilde{D}}_{\mu\nu}\left(p\right) \ \ & {\tilde{G}}_{\mu}\left(p\right)  \   \cr & {\tilde{\Delta}}_{\nu}\left(p\right) \ \ & {\tilde{D}}\left(p\right) \ \cr }e^{-ip\cdot\left(x-y\right)}.
\end{eqnarray}

The explicit form of the propagator can be computed by inverting the matrix operator in Eq. (\ref{gfun}), as follows
\begin{eqnarray}
\label{prop2}
\bordermatrix{&  \cr & {\cal{O}}_{\mu\nu}\left(x\right) \ \ & {\cal{O}}_{\mu}\left(x\right)  \   \cr & -{\cal{O}}_{\nu}\left(x\right) \ \ & -\Box\left(x\right) \ \cr }\bordermatrix{&  \cr & D^{\nu}_{\ \beta}\left(x,y\right) \ \ & G^{\nu}\left(x,y\right)  \   \cr & \Delta_{\beta}\left(x,y\right) \ \ & D\left(x,y\right) \ \cr }=\bordermatrix{&  \cr & \eta_{\mu\beta} \ \ & 0  \   \cr & 0 \ \ & 1 \ \cr } \delta^{3}\left(x-y\right) \ .
\end{eqnarray}

Even if we are able to find explicit forms for the propagators in (\ref{prop2}), the integrals we will encounter when calculating the interaction energies will not be solvable. Therefore, we will have to make an assumption regarding the parameters $(v^{\mu})^2$ and $m^2$, which is to say, either assume $(v^{\mu})^2 \ll m^2$ or $(v^{\mu})^2 \gg m^2$, in order to obtain analytic results. Both choices are possible in principle but, in this work we shall restrict to the case where the anisotropic effects are small. In this case, assuming $(v^{\mu})^2 \ll m^2$ is more natural. We will then assume $(v^{\mu})^2/m^2 \ll 1$ from now on, and calculate the propagators up to the second order of this small parameter. We stress that this is a choice, which could alternatively be justified by simply considering Eq.~\eqref{lagCPToddRd} as our initial theory, disregarding its relation to the higher dimensional SME.

Taking this into consideration, to find explicit forms for the propagators we fix the Feynman gauge $\alpha=1$ and expand up to the order of $(v^{\mu})^2/m^2$, thus obtaining
\begin{eqnarray}
\label{propem}
{\tilde{D}}^{\mu\nu}\left(p\right)&\cong & -\frac{1}{p^{2}-m^{2}}\left(\eta^{\mu\nu}-m^{2}\frac{p^{\mu}p^{\nu}}{p^{4}}+\frac{im}{p^{2}}\epsilon^{\mu\nu\kappa}p_{\kappa}\right)\nonumber\\
&
&+\frac{1}{p^{2}\left(p^{2}-m^{2}\right)^{2}}\left(m^{2}v^{\mu}v^{\nu}+\epsilon^{\mu\alpha\kappa}v_{\alpha}p_{\kappa}\epsilon^{\nu\beta\rho}v_{\beta}p_{\rho}-imv^{\mu}\epsilon^{\nu\beta\rho}v_{\beta}p_{\rho}+imv^{\nu}\epsilon^{\mu\alpha\kappa}v_{\alpha}p_{\kappa}\right)\nonumber\\
&
&-\frac{1}{\left[p^{2}\left(p^{2}-m^{2}\right)\right]^{2}}\Bigl(-m^{2}\left(v\cdot p\right)^{2}\frac{p^{\mu}p^{\nu}}{p^{2}}+m^{2}\left(v\cdot p\right)v^{\mu}p^{\nu}+m^{2}\left(v\cdot p\right)p^{\mu}v^{\nu}\nonumber\\
&
&+im\left(v\cdot p\right)p^{\nu}\epsilon^{\mu\alpha\kappa}v_{\alpha}p_{\kappa}-im\left(v\cdot p\right)p^{\mu}\epsilon^{\nu\beta\rho}v_{\beta}p_{\rho}\Bigr) \ , 
\end{eqnarray}
\begin{eqnarray}
\label{propmix1}
{\tilde{\Delta}}^{\nu}\left(p\right)\cong  -\frac{1}{p^{2}\left(p^{2}-m^{2}\right)}\left(-i\epsilon^{\nu\beta\rho}v_{\beta}p_{\rho}+mv^{\nu}-m\frac{\left(v\cdot p\right)}{p^{2}}p^{\nu}\right) \ , 
\end{eqnarray}
\begin{eqnarray}
\label{propmix2}
{\tilde{G}}^{\mu}\left(p\right)\cong  -\frac{1}{p^{2}\left(p^{2}-m^{2}\right)}\left(i\epsilon^{\mu\beta\rho}v_{\beta}p_{\rho}+mv^{\mu}-m\frac{\left(v\cdot p\right)}{p^{2}}p^{\mu}\right) \ , 
\end{eqnarray}
and
\begin{eqnarray}
\label{propsc}
{\tilde{D}}\left(p\right)\cong  \frac{1}{p^{2}}\left(1+\frac{\left(v\cdot p\right)^{2}}{p^{2}\left(p^{2}-m^{2}\right)}-\frac{v^{2}}{p^{2}-m^{2}}\right) \ .
\end{eqnarray}

We point out that the propagators (\ref{propem}), (\ref{propmix1}), (\ref{propmix2}), (\ref{propsc}) were computed perturbatively up to second order in the background vector $v^{\mu}$. 

For time independent field sources, the functional generator can be written as follows \cite{Zee}
\begin{eqnarray}
\label{fgenEn}
{\cal{Z}}_{(2+1)}=\lim_{T\rightarrow\infty}\exp\left(-iET\right) \ ,
\end{eqnarray}
where $E$ is the vacuum energy of the system and $T$ is a time interval. Comparing Eqs. (\ref{gfunGau}) and (\ref{fgenEn}), we have
\begin{eqnarray}
\label{EnergyS}
E=\frac{1}{2T}\int\int d^{3}x d^{3}y\bordermatrix{&  \cr & J^{\mu}\left(x\right) \  & -J\left(x\right)  \   \cr }\bordermatrix{&  \cr & D_{\mu\nu}\left(x,y\right) \ \ & G_{\mu}\left(x,y\right)  \   \cr & \Delta_{\nu}\left(x,y\right) \ \ & D\left(x,y\right) \ \cr }
\bordermatrix{&  \cr & J^{\nu}\left(y\right) \ \ \cr & -J \left(y\right) \   \cr } \ .
\end{eqnarray}

Substituting (\ref{prop1}) in (\ref{EnergyS}), it can be shown that the energy is given by the sum of the following four parts
\begin{eqnarray}
\label{energyEM}
E_{EM}&=&\frac{1}{2T}\int\frac{d^{3} p}{\left(2\pi\right)^{3}}\int\int d^{3}x \  d^{3}y \ e^{-ip\cdot\left(x-y\right)}J_{\mu}\left(x\right){\tilde{D}}^{\mu\nu}\left(p\right)J_{\nu}\left(y\right) \ ,
\end{eqnarray}
\begin{eqnarray}
\label{energyM1}
E_{M1}&=&-\frac{1}{2T}\int\frac{d^{3} p}{\left(2\pi\right)^{3}}\int\int d^{3}x \  d^{3}y \ e^{-ip\cdot\left(x-y\right)}J\left(x\right){\tilde{\Delta}}^{\nu}\left(p\right)J_{\nu}\left(y\right) \ , 
\end{eqnarray}
\begin{eqnarray}
\label{energyM2}
E_{M2}&=&-\frac{1}{2T}\int\frac{d^{3} p}{\left(2\pi\right)^{3}}\int\int d^{3}x \  d^{3}y \ e^{-ip\cdot\left(x-y\right)}J_{\mu}\left(x\right){\tilde{G}}^{\mu}\left(p\right)J\left(y\right) \ , 
\end{eqnarray}
and
\begin{eqnarray}
\label{energySC}
E_{SC}&=&\frac{1}{2T}\int\frac{d^{3} p}{\left(2\pi\right)^{3}}\int\int d^{3}x \  d^{3}y \ e^{-ip\cdot\left(x-y\right)}J\left(x\right){\tilde{D}}\left(p\right)J\left(y\right) \ ,
\end{eqnarray}
where the labels $EM,M1,M2$, and $SC$ mean the interaction energy in the electromagnetic, mixed and scalar sectors, respectively.

By using Eqs. (\ref{energyEM}), (\ref{energyM1}), (\ref{energyM2}) and (\ref{energySC}), in the next sections we will explore some physical phenomena due to the presence of steady point-like field sources in the proposed Lorentz violating scenario.

\section{POINT-LIKE PLANAR CHARGES }
\label{charges}

In this section we investigate the interaction between two point-like charges for all the sectors of the planar theory (\ref{lagCPToddRd}). The corresponding external sources are given by
\begin{eqnarray}
\label{sourceCC}
J_{\mu}^{CC}\left(x\right)&=& q_{1}\eta_{\mu}^{\ 0}\delta^{2}\left({\bf{x}}-{\bf{a}}_{1}\right)+q_{2}\eta_{\mu}^{\ 0}\delta^{2}\left({\bf{x}}-{\bf{a}}_{2}\right) \ , \\
\label{sourcebarCC}
J^{{\bar{C}}{\bar{C}}}\left(x\right)&=&\sigma_{1}\delta^{2}\left({\bf{x}}-{\bf{a}}_{1}\right)+\sigma_{2}\delta^{2}\left({\bf{x}}-{\bf{a}}_{2}\right) \ ,
\end{eqnarray}
where Eq. (\ref{sourceCC}) stands for two electric planar charges $q_{1}$ and $q_{2}$ placed at positions ${\bf{a}}_{1}$ and ${\bf{a}}_{2}$, and (\ref{sourcebarCC}) for two scalar planar charges $\sigma_{1}$ and $\sigma_{2}$ located at the same positions. The super-indexes $CC$ and ${\bar{C}}{\bar{C}}$ mean that we have a physical system composed by both electric and scalar planar charges, respectively. 

It is important to notice that $q_{1}$ and $q_{2}$ have dimension of electric charge per length. They play the role of what would be electric charges in two dimensional spaces, so we refer to $q_{1}$ and $q_{2}$ as planar electric charges and $\sigma_{1}$ and $\sigma_{2}$ as planar scalar charges.

First, let us consider the electromagnetic sector. Substituting (\ref{propem}) and (\ref{sourceCC}) in (\ref{energyEM}), discarding the self-interaction contributions, we arrive at
\begin{eqnarray}
\label{energyCCEM1}
E_{EM}^{CC}=q_{1}q_{2}\left[\int\frac{d^{2}{\bf{p}}}{\left(2\pi\right)^{2}}\frac{e^{i{\bf{p}}\cdot{\bf{a}}}}{{\bf{p}}^{2}+m^{2}}-\left(m^{2}\left(v^{0}\right)^{2}-\left[\left({\bf{v}}\times{\hat{z}}\right)\cdot{\bf{\nabla}}_{\bf{a}}\right]^{2}\right)\int\frac{d^{2}{\bf{p}}}{\left(2\pi\right)^{2}}\frac{e^{i{\bf{p}}\cdot{\bf{a}}}}{{\bf{p}}^{2}\left({\bf{p}}^{2}+m^{2}\right)^{2}}\right] \ ,
\end{eqnarray}
with ${\bf{a}}={\bf{a}}_{1}-{\bf{a}}_{2}$ standing for the distance between the sources, and ${\bf{\nabla}}_{\bf{a}}=\left(\frac{\partial}{\partial a^{1}},\frac{\partial}{\partial a^{2}}\right)$.

Performing the integrals and acting with the differential operator, we obtain
\begin{eqnarray}
\label{energyCCEM2}
E_{EM}^{CC}&=&\frac{q_{1}q_{2}}{2\pi}\Biggl[K_{0}\left(ma\right)+\frac{\left(v^{0}\right)^{2}}{m^{2}}\left(\ln\left(\frac{a}{a_{0}}\right)+K_{0}\left(ma\right)+\frac{1}{2}\left(ma\right)K_{1}\left(ma\right)\right)\nonumber\\
&
&-\frac{\left[\left({\bf{v}}\times{\hat{z}}\right)\cdot{\bf{a}}\right]^{2}}{\left(ma\right)^{2}}\left(\frac{1}{2}\left(ma\right)K_{1}\left(ma\right)+K_{2}\left(ma\right)\right)+\frac{{\bf{v}}^{2}}{2m^{2}}K_{2}\left(ma\right)\Biggr] \ ,
\end{eqnarray}
where $K$ stands for the $K$-Bessel function \cite{Arfken}, $a=\mid{\bf{a}}\mid$, and $a_{0}$ is an arbitrary constant with dimension of length, which was introduced in order to make the argument of the logarithmic function dimensionless. We also assumed that $\frac{v^{2}}{m^{2}}\ll \left(ma\right)$, so we have discarded contributions proportional to $\frac{{\bf{v}}^{2}}{m^{4}a^{2}}$ and $\frac{{\bf{v}}^{2}}{m^{3}{{a}}}$. 

The expression (\ref{energyCCEM2}) is the interaction energy between two electric planar charges up to order of $\frac{v^{2}}{m^{2}}$. The first contribution between brackets on the right hand side of the Eq. (\ref{energyCCEM2}) stands for the standard result obtained in Maxwell-Chern-Simons electrodynamics \cite{sourcesMCS}, while the remaining terms are Lorentz violation corrections induced by the background vector $v^{\mu}$.

\begin{figure}[!h]
\centering \includegraphics[scale=0.20]{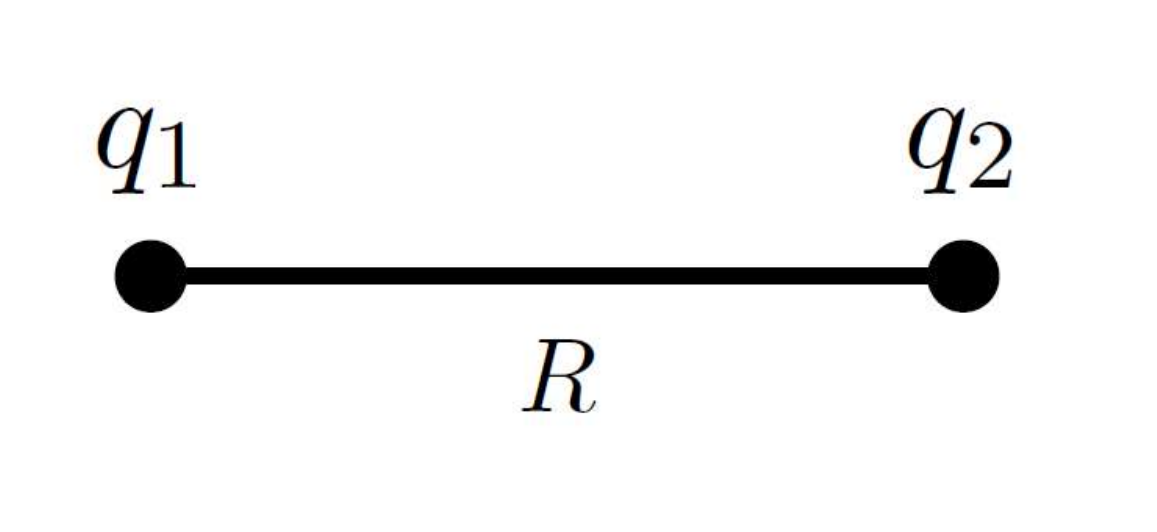} \caption{Physical system representing the electromagnetic sector. In this figure, the line can be interpreted both as the spatial separation of the two planar charges, as well as the fact that they interact directly via the electrogmagnetic field.}
\label{figura1}
\end{figure}

Taking the distance between the two electric planar charges as fixed ${\bf{a}}={\bf{R}}$, the energy (\ref{energyCCEM2})  leads to an interesting effect due to the orientation of  ${\bf{a}}$ with respect to the background vector, namely an spontaneous torque on this system. In Fig. (\ref{figura1}), we have a plot representing such a physical system. Defining as $\theta\in[0,2\pi)$ the angle between ${\bf{v}}$ and ${\bf{a}}$, the energy becomes
\begin{eqnarray}
\label{energyCCEM3}
E_{EM}^{CC}\left(\theta\right)&=&\frac{q_{1}q_{2}}{2\pi}\Biggl[K_{0}\left(mR\right)+\frac{\left(v^{0}\right)^{2}}{m^{2}}\left(\ln\left(\frac{R}{R_{0}}\right)+K_{0}\left(mR\right)+\frac{1}{2}\left(mR\right)K_{1}\left(mR\right)\right)\nonumber\\
&
&-\frac{{\bf{v}}^{2}}{m^{2}}\sin^{2}\left(\theta\right)\left(\frac{1}{2}\left(mR\right)K_{1}\left(mR\right)+K_{2}\left(mR\right)\right)+\frac{{\bf{v}}^{2}}{2m^{2}}K_{2}\left(mR\right)\Biggr] \ ,
\end{eqnarray}
where $R=\mid{\bf{R}}\mid$.  The spontaneous torque can be computed as follows 
\begin{eqnarray}
\label{torqueemCC}
\tau_{EM}^{CC}&=&-\frac{\partial E_{EM}^{CC}\left(\theta\right)}{\partial\theta}\nonumber\\
&=&\frac{q_{1}q_{2}}{2\pi}\frac{{\bf{v}}^{2}}{m^{2}}\sin\left(2\theta\right)\left(\frac{1}{2}\left(mR\right)K_{1}\left(mR\right)+K_{2}\left(mR\right)\right) \ .
\end{eqnarray}

\begin{figure}[!h]
\centering \includegraphics[scale=0.4]{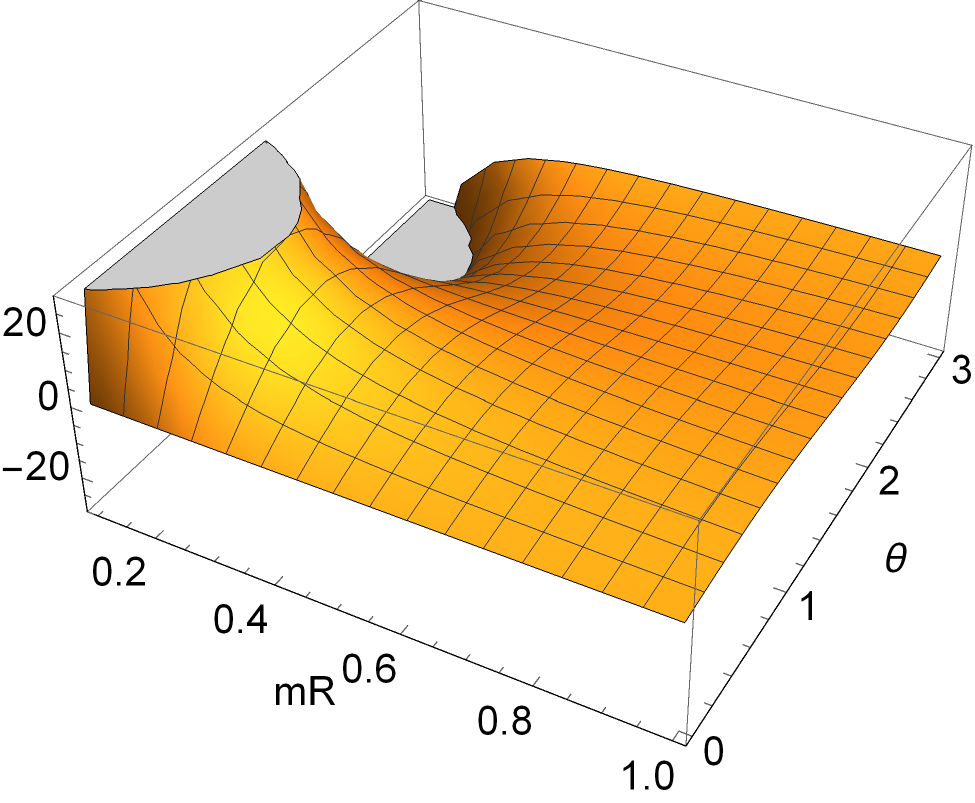} \caption{The torque given in Eq. (\ref{torqueemCC}), multiplied by $\frac{(2\pi)m^{2}}{q_{1}q_{2}{\bf{v}}^{2}}$.}
\label{figura2}
\end{figure}

The torque (\ref{torqueemCC}) does not occur in the usual Maxwell-Chern-Simons electrodynamics so, it is an exclusive Lorentz violating planar effect. If $\mid{\bf{v}}\mid=0$ or $\theta=0,\pi/2,\pi$ this effect is absent, for $\theta=\pi/4$ the torque reaches the maximum intensity. For the special situation where we have two opposite electric planar charges $q_{1}=-q_{2}=q$, the expression (\ref{torqueemCC}) gives an spontaneous torque on an electric dipole. In Fig. (\ref{figura2}), it is shown the general behavior of the torque (\ref{torqueemCC}) as a function of $mR$ and $\theta$.

Another physical phenomenon that can be obtained from Eq. (\ref{energyCCEM2}) is an interaction force between the planar charges, which is given by
\begin{eqnarray}
\label{forCCem}
{\bf{F}}_{EM}^{CC}&=&-{\bf{\nabla}}_{{\bf{a}}}E_{EM}^{CC}\nonumber\\
&=&-\frac{mq_{1}q_{2}}{2\pi}\Biggl\{\Biggl[-K_{1}\left(ma\right)+\frac{\left(v^{0}\right)^{2}}{m^{2}}\Biggl(\frac{1}{ma}-\frac{1}{2}\left(ma\right)K_{2}\left(ma\right)\Biggr)\nonumber\\
&
&+\frac{\left[\left({\bf{v}}\times{\hat{z}}\right)\cdot{\bf{a}}\right]^{2}}{\left(ma\right)^{2}}\Biggl(\frac{[8+\left(ma\right)^{2}]}{2\left(ma\right)}K_{0}\left(ma\right)+\frac{2[4+\left(ma\right)^{2}]}{\left(ma\right)^{2}}K_{1}\left(ma\right)\Biggr)\nonumber\\
&
&-\frac{{\bf{v}}^{2}}{4m^{2}}\left[K_{1}\left(ma\right)+K_{3}\left(ma\right)\right]\Biggr]{\hat{a}}\nonumber\\
&
&-\frac{2\left[\left({\bf{v}}\times{\hat{z}}\right)\cdot{\bf{a}}\right]}{ma}\Biggl(\frac{1}{2}K_{1}\left(ma\right)+\frac{K_{2}\left(ma\right)}{ma}\Biggr)\Bigl(\frac{{\bf{v}}}{m}\times{\hat{z}}\Bigr)\Biggr\} \ ,
\end{eqnarray}
where ${\hat{a}}$ is the unit vector pointing on the direction of ${\bf{a}}$.

Notice that Eq. (\ref{forCCem}) is a perturbative result up to order of $\frac{v^{2}}{m^{2}}$, which shows in a more explicit way the anisotropies generated by $v^{\mu}$. In the absence of the background vector, we recover the standard result computed in Maxwell-Chern-Simons electrodynamics \cite{sourcesMCS}.

For the scalar sector, we substitute Eqs. (\ref{propsc}) and (\ref{sourcebarCC}) in (\ref{energySC}), and then we discard the self-interaction contributions, obtaining
\begin{eqnarray}
\label{enerscbar}
E_{SC}^{{\bar{C}}{\bar{C}}}=-\sigma_{1}\sigma_{2}\left[\int\frac{d^{2}{\bf{p}}}{\left(2\pi\right)^{2}}\frac{e^{i{\bf{p}}\cdot{\bf{a}}}}{{\bf{p}}^{2}}-\left({\bf{v}}\cdot{\bf{\nabla}}_{{\bf{a}}}\right)^{2}\int\frac{d^{2}{\bf{p}}}{\left(2\pi\right)^{2}}\frac{e^{i{\bf{p}}\cdot{\bf{a}}}}{{\bf{p}}^{4}\left({\bf{p}}^{2}+m^{2}\right)}+v^{2}\int\frac{d^{2}{\bf{p}}}{\left(2\pi\right)^{2}}\frac{e^{i{\bf{p}}\cdot{\bf{a}}}}{{\bf{p}}^{2}\left({\bf{p}}^{2}+m^{2}\right)}\right].
\end{eqnarray}
Carrying out the calculations, we end up with
\begin{eqnarray}
\label{enerscbar2}
E_{SC}^{{\bar{C}}{\bar{C}}}&=&\frac{\sigma_{1}\sigma_{2}}{2\pi}\Biggl[\ln\left(\frac{a}{a_{0}}\right)+\frac{\left(v^{0}\right)^{2}}{m^{2}}\Biggl(\ln\left(\frac{a}{a_{0}}\right)+K_{0}\left(ma\right)\Biggr)\nonumber\\
&
&-\frac{{\bf{v}}^{2}}{2m^{2}}\Biggl(\ln\left(\frac{a}{a_{0}}\right)+K_{0}\left(ma\right)+K_{2}\left(ma\right)\Biggr)+\frac{\left({\bf{v}}\cdot{\bf{a}}\right)^{2}}{\left(ma\right)^{2}}\Biggl(\frac{1}{2}+K_{2}\left(ma\right)\Biggr)\Biggr] \, .
\end{eqnarray}

Equation (\ref{enerscbar2}) stands for the interaction energy between two scalar planar charges up to order of $\frac{v^{2}}{m^{2}}$. The first term on the right hand side of the Eq. (\ref{enerscbar2}) correspond to the result of the usual massless Klein-Gordon field theory \cite{FAB}, the additional terms are corrections due to the Lorentz symmetry breaking.

\begin{figure}[!h]
\centering \includegraphics[scale=0.20]{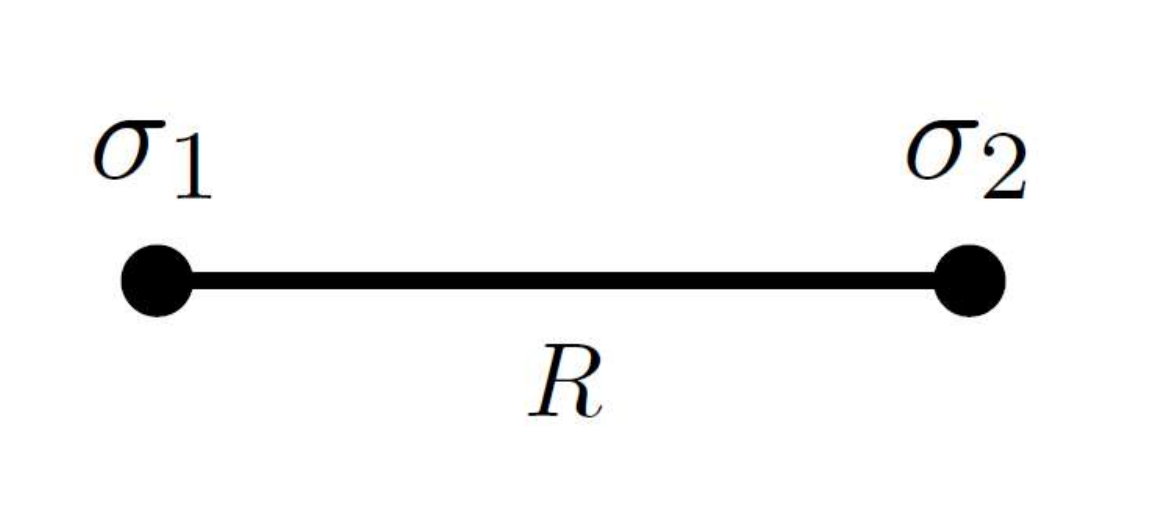} \caption{Representation of the physical system studied in the scalar sector. In this figure, the line can be interpreted both as the spatial separation of the two planar charges, as well as the fact that they interact directly via the scalar field.}
\label{figura3}
\end{figure}

Similarly to what was done previously for electric planar charges, when we fix the distance between the two scalar planar charges, the expression (\ref{enerscbar2}) leads to an spontaneous torque on the system  represented in Fig. (\ref{figura3}), as follows
\begin{eqnarray}
\label{torqueSC}
\tau_{SC}^{{\bar{C}}{\bar{C}}}=\frac{\sigma_{1}\sigma_{2}}{2\pi}\frac{{\bf{v}}^{2}}{m^{2}}\sin\left(2\theta\right)\left(\frac{1}{2}+K_{2}\left(mR\right)\right)  \, .
\end{eqnarray}

We notice that Eq. (\ref{torqueSC}) is another planar physical phenomenon solely due to the Lorentz violation, which exhibits a behavior similar to the one observed in the electromagnetic sector. For the case where we have $\sigma_{1}=-\sigma_{2}=\sigma$, expression (\ref{torqueSC}) becomes the spontaneous torque on a classical scalar dipole.

From Eq. (\ref{enerscbar2}), the interaction force in the scalar sector up to order of $\frac{v^{2}}{m^{2}}$ reads
\begin{eqnarray}
\label{ForceSC}
{\bf{F}}_{SC}^{{\bar{C}}{\bar{C}}}&=&-\frac{\sigma_{1}\sigma_{2}}{2\pi a}\Biggl\{\Biggl[1+\frac{\left(v^{0}\right)^{2}}{m^{2}}\Bigl(1-\left(ma\right)K_{1}\left(ma\right)\Bigr)-\frac{{\bf{v}}^{2}}{m^{2}}\Biggl(\frac{1}{2}-K_{0}\left(ma\right)-\frac{[2+\left(ma\right)^{2}]}{ma}K_{1}\left(ma\right)\Biggr)\nonumber\\
&
&-\frac{\left({\bf{v}}\cdot{\bf{a}}\right)^{2}}{\left(ma\right)^{2}}\Biggl(1+\left(ma\right)K_{3}\left(ma\right)\Biggr)\Biggr]{\hat{a}}+\frac{\left({\bf{v}}\cdot{\bf{a}}\right)}{ma}\Bigl(1+2K_{2}\left(ma\right)\Bigr)\frac{{\bf{v}}}{m}\Biggr\} .
\end{eqnarray}
For the situation where ${\bf{v}}$ is perpendicular to ${\bf{a}}$, the interaction force in (\ref{ForceSC})   just points on the direction of ${\bf{a}}$. As expected, if $v^{\mu}$ vanishes, Eq. (\ref{ForceSC}) reduces to the result obtained in standard massless scalar field theory \cite{FAB}.

Finally, for the mixed sector, by using  Eqs. (\ref{propmix1}), (\ref{propmix2}), (\ref{sourceCC}), (\ref{sourcebarCC}), (\ref{energyM1}), (\ref{energyM2}), one can show that
\begin{eqnarray}
E_{M1}^{C{\bar{C}}}=E_{M2}^{C{\bar{C}}}&=&-\frac{1}{2}\Biggl[\sigma_{1}q_{2}\Bigl(\left[\left({\bf{v}}\times{\hat{z}}\right)\cdot{\bf{\nabla}}_{\bf{a}}\right]+mv^{0}\Bigr)\nonumber\\
&
&+q_{1}\sigma_{2}\Bigl(-\left[\left({\bf{v}}\times{\hat{z}}\right)\cdot{\bf{\nabla}}_{\bf{a}}\right]+mv^{0}\Bigr)\Biggr]\int\frac{d^{2}{\bf{p}}}{\left(2\pi\right)^{2}}\frac{e^{i{\bf{p}}\cdot{\bf{a}}}}{{\bf{p}}^{2}\left({\bf{p}}^{2}+m^{2}\right)} \ ,
 \end{eqnarray}
leading to
\begin{eqnarray}
\label{EM1EM21}
E_{M1}^{C{\bar{C}}}=E_{M2}^{C{\bar{C}}}&=&-\frac{1}{4\pi}\Biggl[\left(\sigma_{1}q_{2}+q_{1}\sigma_{2}\right)\frac{v^{0}}{m}\Biggl(\ln\left(\frac{a}{a_{0}}\right)+K_{0}\left(ma\right)\Biggr)\nonumber\\
&
&-\left(\sigma_{1}q_{2}-q_{1}\sigma_{2}\right)\frac{\left[\left({\bf{v}}\times{\hat{z}}\right)\cdot{\bf{a}}\right]}{ma}\Biggl(K_{1}\left(ma\right)-\frac{1}{ma}\Biggr)\Biggr] \ ,
\end{eqnarray}
where the label $C{\bar{C}}$ stands for interactions involving electric and scalar planar charges.

The expression (\ref{EM1EM21}) shows that electric planar charges can interact with scalar planar charges in $2+1$ dimensions due to the presence of the background vector, since this interaction disappears if $v^{\mu}=0$.
In the setup where ${\bf{v}}$ is parallel to ${\bf{a}}$, the energy depends only on ${v^{0}}$.

\begin{figure}[!h]
\centering \includegraphics[scale=0.20]{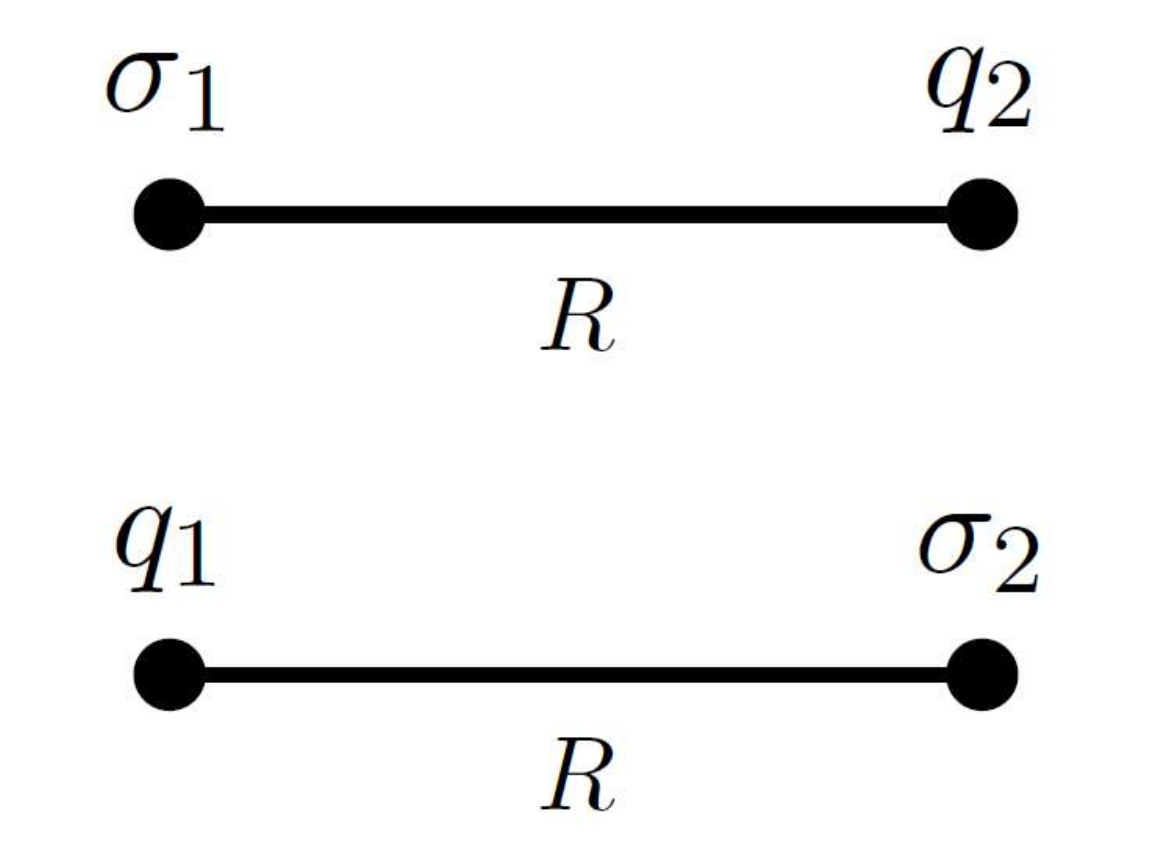} \caption{Representation of the physical system studied the mixed sector. Here, scalar and electromagnetic planar charges are spatially coincident at two points separated by the distance $R$. In the figure, the lines represent the direct interactions between these different types of sources.}
\label{figura4}
\end{figure}

In the mixed sector, we also have the appearance of a spontaneous torque, now on the system shown in Fig. (\ref{figura4}). Such torque can be obtained from Eq. (\ref{EM1EM21}), giving the following result
\begin{eqnarray}
\label{torqueM1M2}
\tau_{M}^{C{\bar{C}}}&=&\tau_{M1}^{C{\bar{C}}}+\tau_{M2}^{C{\bar{C}}}\nonumber\\
&=&\frac{\left(\sigma_{1}q_{2}-q_{1}\sigma_{2}\right)}{2\pi}\frac{\mid{\bf{v}}\mid}{m}\cos\left(\theta\right)\Biggl(K_{1}\left(mR\right)-\frac{1}{mR}\Biggr) \ .
\end{eqnarray}

\begin{figure}[!h]
\centering \includegraphics[scale=0.4]{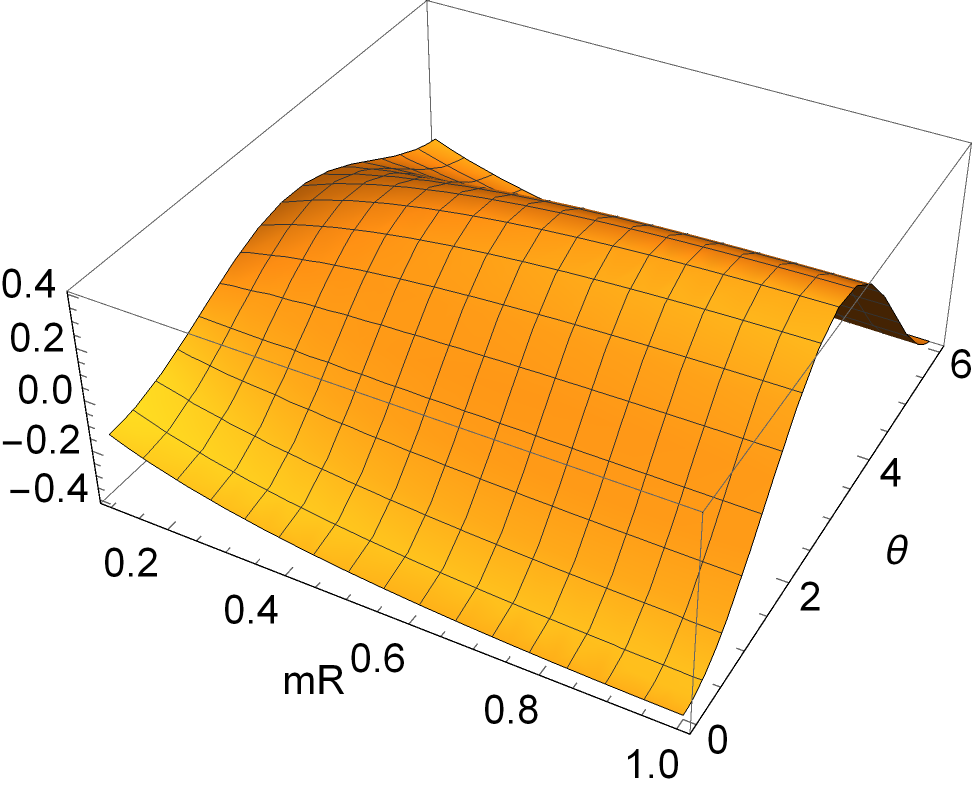} \caption{The torque given in Eq. (\ref{torqueM1M2}), multiplied by $\frac{(2\pi)m}{\left(\sigma_{1}q_{2}-q_{1}\sigma_{2}\right)\mid{\bf{v}}\mid}$.}
\label{figura5}
\end{figure}

As expected, this torque vanishes when ${|\bf{v}|}=0$, as well as for the special configurations with $\theta=\pi/2$ or $\sigma_{1}q_{2}=q_{1}\sigma_{2}$. In Fig. (\ref{figura5}), we have a plot for the torque (\ref{torqueM1M2}) multiplied by $\frac{(2\pi)m}{\left(\sigma_{1}q_{2}-q_{1}\sigma_{2}\right)\mid{\bf{v}}\mid}$. If $\sigma_{1}=q_{2}=0, q_{1}=q, \sigma_{2}=-\sigma, \mid q\mid=\mid\sigma\mid, q>0, \sigma>0$, Eq.~(\ref{torqueM1M2}) leads to a spontaneous torque on a mixed dipole. A similar situation occurs if $\sigma_{2}=q_{1}=0, q_{2}=-q, \sigma_{1}=\sigma, \mid q\mid=\mid\sigma\mid, q>0, \sigma>0$. It is important to highlight that the presence of a mixed dipole implies the non-existence of the purely electromagnetic and scalar interactions that we have studied previously, therefore the torque in~\eqref{torqueM1M2} is the only effect present in this case.

\begin{figure}[!h]
\centering \includegraphics[scale=0.20]{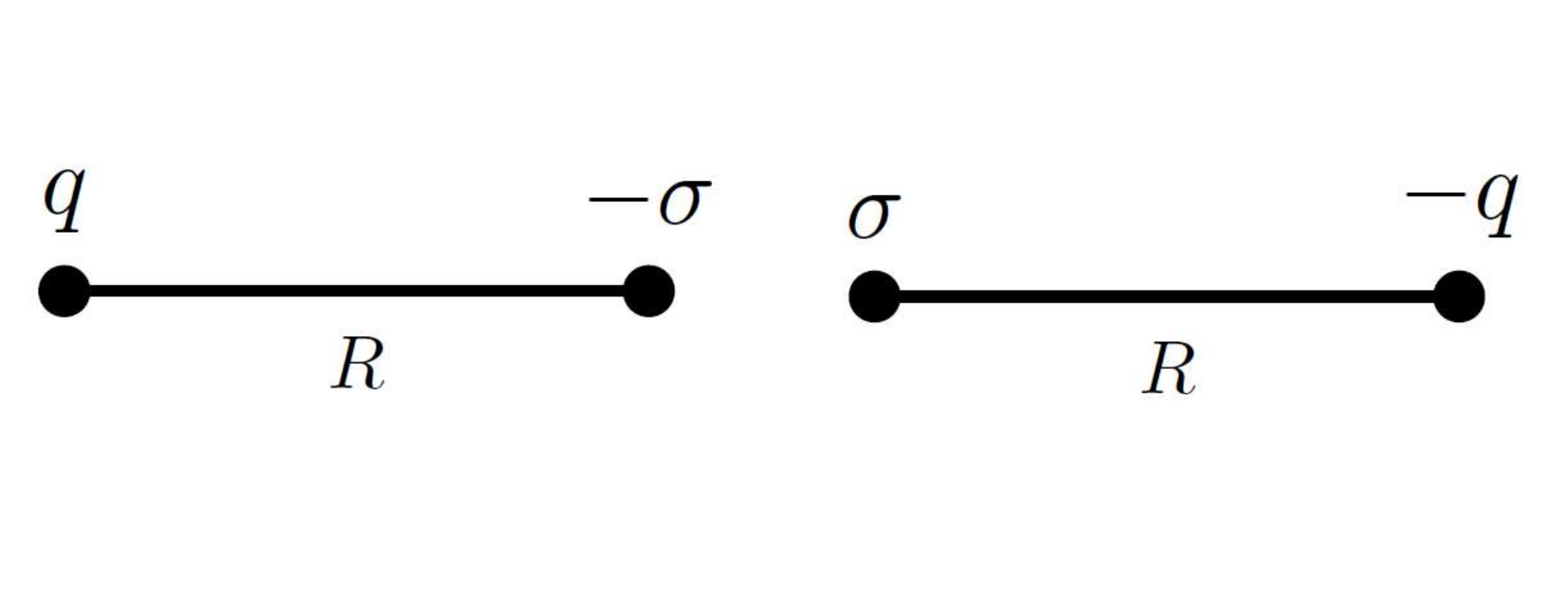} \caption{Mixed dipoles, as discussed in the paragraph following Eq.\,\eqref{torqueM1M2}. }
\label{figura6}
\end{figure}

The interaction force in the mixed sector is given by,
\begin{eqnarray}
\label{ForceM}
{\bf{F}}_{M}^{C{\bar{C}}}&=&{\bf{F}}_{M1}^{C{\bar{C}}}+{\bf{F}}_{M2}^{C{\bar{C}}}\nonumber\\
&=&\frac{m}{2\pi}\Biggl\{\Biggl[\left(\sigma_{1}q_{2}+q_{1}\sigma_{2}\right)\frac{v^{0}}{m}\Biggl(\frac{1}{ma}-K_{1}\left(ma\right)\Biggr)-\left(\sigma_{1}q_{2}-q_{1}\sigma_{2}\right)\frac{\left[\left({\bf{v}}\times{\hat{z}}\right)\cdot{\bf{a}}\right]}{ma}\nonumber\\
&
&\times\Biggl(\frac{2}{\left(ma\right)^{2}}-K_{2}\left(ma\right)\Biggr)\Biggr]{\hat{a}}-\frac{\left(\sigma_{1}q_{2}-q_{1}\sigma_{2}\right)}{ma}\Biggl(K_{1}\left(ma\right)-\frac{1}{ma}\Biggr)\Bigl(\frac{{\bf{v}}}{m}\times{\hat{z}}\Bigr)\Biggr\} \ ,
\end{eqnarray}
which is a new planar effect that occurs only due to the presence of the background vector.

As final comment, we point out that to obtain the interaction energy, the torque, and the interaction force for the whole theory, it is enough to add the corresponding contributions obtained separately for each sector.

\section{DIRAC POINTS }
\label{DIRAC}

In the present section we obtain some physical phenomena due to the presence of Dirac points \cite{sourcesMCS,sourcesHMCS}. A Dirac point is the source for vortex field solutions, which can be obtained via dimensional reduction of the external source produced by the Dirac string in $(3+1)$ dimensions \cite{ce1,ce2,nmsources}. Besides, in the standard Maxwell-Chern-Simons electrodynamics, taking into account the interaction between a Dirac point and an electric planar charge, as well as the interaction between two Dirac points, we notice that the Dirac point behaves like an electric planar charge (in the sense of leading to the same interaction energy, interchangeably) \cite{sourcesMCS}. It is also interesting to investigate whether this feature remains valid in a Lorentz violating scenario.

We start by considering the interaction between point-like planar charges and a Dirac point for all the sectors of the $(2+1)$-dimensional theory (\ref{lagCPToddRd}). The scalar field source is given by the Eq. (\ref{sourcebarCC}) and the electromagnetic one reads
\begin{eqnarray}
\label{sourceCD}
J^{CD}_{\mu}(x)=q\eta_{\ \mu}^{0}\delta^{2}\left({\bf x}-{\bf a}_ {1}\right)+
J_{\mu(D)}\left({\bf x}\right) \ ,
\end{eqnarray}
where the first contribution corresponds to an electric planar charge placed at position ${{\bf {a}}_{1}}$, and the second one to the Dirac point, which is labelled by $D$. The Dirac point is located at the position 
${{\bf {a}}_{2}}$ with a magnetic flux $\Phi$. Such an external source is described by \cite{sourcesMCS,sourcesHMCS}
\begin{eqnarray}
\label{sourceD}
J^{\mu}_{(D)}\left({\bf x}\right)=-2\pi i\Phi\int\frac{d^{3}p}{\left(2\pi\right)^{3}} \ 
\delta\left(p^{0}\right)\epsilon^{0\mu\alpha}p_{\alpha} \ 
e^{-i p\cdot x}e^{-i {\bf{p}}\cdot{\bf {a}_{2}}} \ .
\end{eqnarray}

For the electromagnetic sector, we must substitute (\ref{propem}), (\ref{sourceCD}), (\ref{sourceD}) in Eq. (\ref{energyEM}) and discard the self-interaction contributions, obtaining
\begin{eqnarray}
\label{energyEMCD1}
E_{EM}^{CD}&=&mq\Phi\Biggl[\int\frac{d^{2}{\bf{p}}}{\left(2\pi\right)^{2}}\frac{e^{i{\bf{p}}\cdot{\bf{a}}}}{{\bf{p}}^{2}+m^{2}}+\Bigl(-mv^{0}\left[\left({\bf{v}}\times{\hat{z}}\right)\cdot{\bf{\nabla}}_{\bf{a}}\right]+\left[\left({\bf{v}}\times{\hat{z}}\right)\cdot{\bf{\nabla}}_{\bf{a}}\right]^{2}\Bigr)\nonumber\\
&
&\times\int\frac{d^{2}{\bf{p}}}{\left(2\pi\right)^{2}}\frac{e^{i{\bf{p}}\cdot{\bf{a}}}}{{\bf{p}}^{2}\left({\bf{p}}^{2}+m^{2}\right)^{2}}
+\Biggl(\left(v^{0}\right)^{2}-\frac{v^{0}}{m}\left[\left({\bf{v}}\times{\hat{z}}\right)\cdot{\bf{\nabla}}_{\bf{a}}\right]\Biggr)\int\frac{d^{2}{\bf{p}}}{\left(2\pi\right)^{2}}\frac{e^{i{\bf{p}}\cdot{\bf{a}}}}{\left({\bf{p}}^{2}+m^{2}\right)^{2}}\Biggr] \ ,
\end{eqnarray}
and, finally,
\begin{eqnarray}
\label{energyEMCD2}
E_{EM}^{CD}&=&\frac{mq\Phi}{2\pi}\Biggl[K_{0}\left(ma\right)-\frac{v^{0}}{m}\frac{\left[\left({\bf{v}}\times{\hat{z}}\right)\cdot{\bf{a}}\right]}{ma}\Biggl(K_{1}\left(ma\right)-\frac{1}{ma}\Biggr)+\frac{\left(v^{0}\right)^{2}}{2m^{2}}\left(ma\right)K_{1}\left(ma\right)\nonumber\\
&
&-\frac{\left[\left({\bf{v}}\times{\hat{z}}\right)\cdot{\bf{a}}\right]^{2}}{\left(ma\right)^{2}}\left(\frac{1}{2}\left(ma\right)K_{1}\left(ma\right)+K_{2}\left(ma\right)\right)+\frac{{\bf{v}}^{2}}{2m^{2}}K_{2}\left(ma\right)\Biggr] \ .
\end{eqnarray}

The first contribution between brackets on the right hand side stands for the interaction energy between a point-like planar charge and a Dirac point obtained in the standard Maxwell-Chern-Simons electrodynamics \cite{sourcesMCS}, the remaining terms are Lorentz symmetry breaking corrections up to order of $\frac{(v^{\mu})^{2}}{m^{2}}$. 

We notice that in the setup where $v^{\mu}=\left(0,{\bf{v}}\right)$ the Dirac point behaves as an electric planar charge when it interacts with another electric planar charge.  In order to check this fact, in Eq. (\ref{energyEMCD2}) we must identify $q_{1}=q , q_{2}=m\Phi$ and take $v^{0}=0$ so, we recover the interaction energy between two electric planar charges obtained in Eq. (\ref{energyCCEM2}).

\begin{figure}[!h]
\centering \includegraphics[scale=0.20]{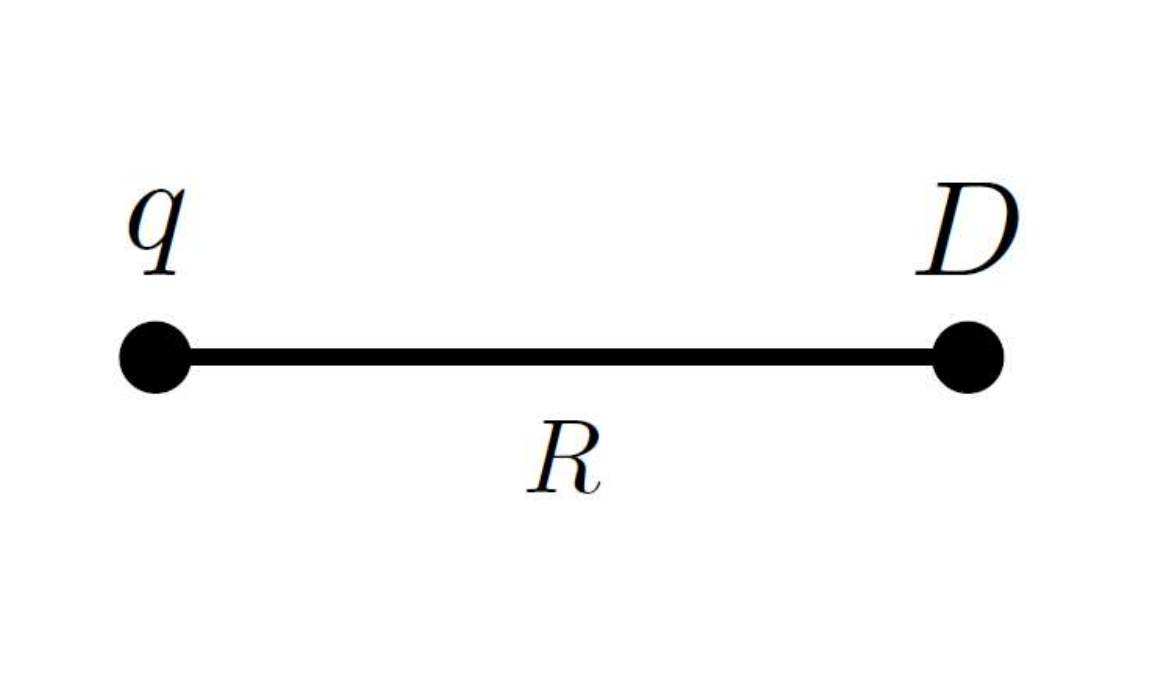} \caption{Electric planar charge and Dirac point separated by a fixed distance $R$. The line is to be interpreted as in Fig.~\ref{figura1}.}
\label{figura7}
\end{figure}

Proceeding in a similar way to what was done in the previous section, the energy (\ref{energyEMCD2}) provides a spontaneous torque on the system that is represented in Fig. (\ref{figura7}), given by
\begin{eqnarray}
\label{torqueEMCD}
\tau_{EM}^{CD}&=&-\frac{mq\Phi}{2\pi}\Biggl[\frac{v^{0}\mid{\bf{v}}\mid}{m^{2}}\cos\left(\theta\right)\Biggl(K_{1}\left(mR\right)-\frac{1}{mR}\Biggr)\nonumber\\
&
&-\frac{{\bf{v}}^{2}}{m^{2}}\sin\left(2\theta\right)\left(\frac{1}{2}\left(mR\right)K_{1}\left(mR\right)+K_{2}\left(mR\right)\right)\Biggr] \ .
\end{eqnarray}

The expression (\ref{torqueEMCD}) is a new physical phenomenon that arises due to the anisotropy created by the presence of the background vector. In the special situation where $v^{0}=0$, the torque exhibits a behavior similar to that one described in the previous section for the electromagnetic sector. Additionally, for $\theta=0,\pi$ the torque becomes proportional to $v^{0}\mid{\bf{v}}\mid$. On the other hand, if $v^{0}=0$ and $\mid q\mid=\mid m\Phi\mid$ with $q>0,\Phi<0$ or $q<0,\Phi>0$, we have a configuration similar to an electric dipole.

From Eqs. (\ref{propmix1}), (\ref{propmix2}), (\ref{sourcebarCC}), (\ref{sourceCD}), (\ref{sourceD}), (\ref{energyM1}), (\ref{energyM2}), it can be shown that the interaction energy for the mixed sector reads
\begin{eqnarray}
\label{energyCCD}
E_{M1}^{CD{\bar{C}}}=E_{M2}^{CD{\bar{C}}}&=&-\frac{1}{4\pi}\Biggl\{\frac{v^{0}}{m}\Biggl[q\sigma_{2}\Biggl(\ln\left(\frac{a}{a_{0}}\right)+K_{0}\left(ma\right)\Biggr)+\left(m\sigma_{1}\Phi\right)K_{0}\left(ma\right)\Biggr]\nonumber\\
&
&-\left(m\sigma_{1}\Phi -q\sigma_{2}\right)\frac{\left[\left({\bf{v}}\times{\hat{z}}\right)\cdot{\bf{a}}\right]}{ma}\Biggl(K_{1}\left(ma\right)-\frac{1}{ma}\Biggr)\Biggr\} \ .
\end{eqnarray}

Equation (\ref{energyCCD}) shows once again that the anisotropy generated by the Lorentz symmetry breaking allows interactions involving both the electromagnetic and scalar sources. The above expression  exhibits two contributions, the first one arises due to the interaction between both electric and scalar planar charges and the another one comes from the interaction of a scalar planar charge with a Dirac point, so we labelled this interaction by the super-index $CD{\bar{C}}$. Considering $v^{0}=0$, the interaction energy (\ref{energyCCD}) disappears if $m\sigma_{1}\Phi = q\sigma_{2}$ or ${\bf{v}}$ is parallel to ${\bf{a}}$.

It is important to highlight that in the setup where $v^{\mu}=\left(0,{\bf{v}}\right)$ and $q_{1}=q,q_{2}=m\Phi$ the expression (\ref{energyCCD})  becomes equivalent to Eq. (\ref{EM1EM21}) therefore, we conclude that the Dirac point also behaves like an electric planar charge when it interacts with a scalar planar charge.

\begin{figure}[!h]
\centering \includegraphics[scale=0.20]{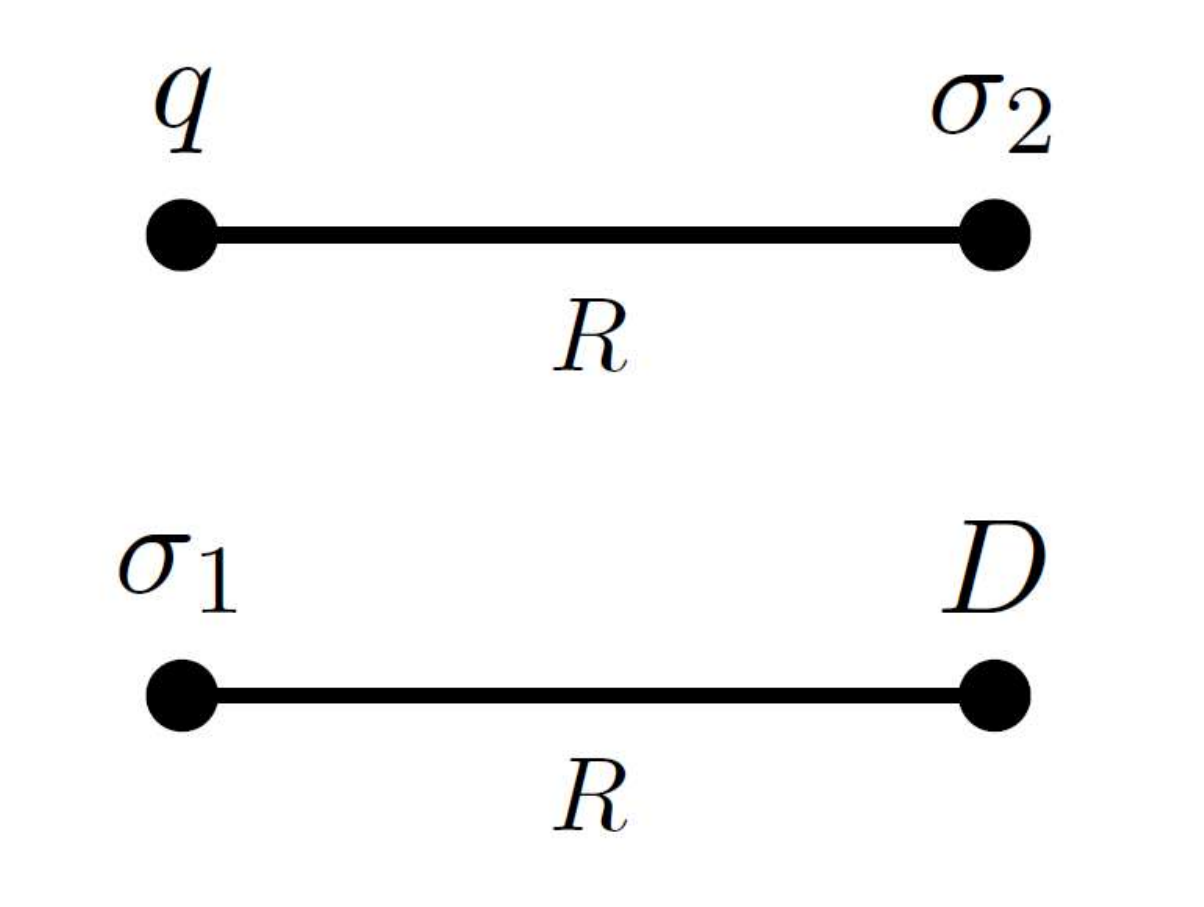} \caption{Representation of the physical system described by the mixed sector. In the figure, the lines represent the direct interactions between these different types of sources, while $q$ and $\sigma_1$ coincide at one spatial location, and $\sigma_2$ and the Dirac point coincide at another location separated by the distance $R$.}
\label{figure8}
\end{figure}

For the physical system indicated in Fig. (\ref{figure8}), the interaction energy (\ref{energyCCD})
leads to an spontaneous torque on this system, given by
\begin{eqnarray}
\label{torqueMCCD}
\tau_{M}^{CD{\bar{C}}}&=&\tau_{M1}^{CD{\bar{C}}}+\tau_{M2}^{CD{\bar{C}}}\nonumber\\
&=&\frac{\left(m\sigma_{1}\Phi -q\sigma_{2}\right)}{2\pi}\frac{\mid{\bf{v}}\mid}{m}\cos\left(\theta\right)\Biggl(K_{1}\left(mR\right)-\frac{1}{mR}\Biggr) \ ,
\end{eqnarray}
which has a similar behavior to that one shown in Fig. (\ref{figura5}). 
Not surprisingly, if $\mid{\bf{v}}\mid=0$ this effect vanishes, but it is interesting to notice that the same happens for the specific configuration $m\sigma_{1}\Phi = q\sigma_{2}$. Also, for $q=\sigma_{2}=0,\mid\sigma_{1}\mid=\mid m\Phi\mid$, if $\sigma_{1}>0,\Phi<0$ or $\sigma_{1}<0,\Phi>0$, we have a structure similar to a mixed dipole. 

For the scalar sector, the results are the same as those obtained in Sect. \ref{charges}. From  Eqs. (\ref{energyEMCD2}) and (\ref{energyCCD}), we can also  obtain the interaction forces for both the electromagnetic and mixed sectors, respectively. 

In the last example, we explore the interactions involving two Dirac points and scalar planar charges. For the scalar sector we consider the source given in Eq. (\ref{sourcebarCC}), while for the electromagnetic sector, we choose a coordinate system where the first Dirac point $(D_{1})$ is placed at the position ${\bf{a}}_{1}$, with magnetic flux $\Phi_{1}$, and the second one $(D_{2})$, with flux $\Phi_{2}$, is concentrated at the position ${\bf{a}}_{2}$. The external source for this setup reads
\begin{eqnarray}
\label{sourceDDEM}
J_{\mu}^{DD}\left({x}\right)=J_{\mu(D_{1})}\left({\bf x}\right)+J_{\mu(D_{2})}\left({\bf x}\right) \ ,
\end{eqnarray}
with
\begin{eqnarray}
\label{sourceD2}
J^{\mu}_{(D_{1})}\left({\bf x}\right)=-2\pi i\Phi_{1}\int\frac{d^{3}p}{\left(2\pi\right)^{3}} \ 
\delta\left(p^{0}\right)\epsilon^{0\mu\alpha}p_{\alpha} \ 
e^{-i p\cdot x}e^{-i {\bf{p}}\cdot{\bf {a}_{1}}}\ ,
\end{eqnarray}
and similarly for $D_{2}$.

For the electromagnetic sector of the planar theory, following the same steps used previously, we arrive at
\begin{eqnarray}
\label{energyEMDD}
E_{EM}^{DD}&=&\frac{m^{2}\Phi_{1}\Phi_{2}}{2\pi}\Biggl[K_{0}\left(ma\right)+\frac{\left(v^{0}\right)^{2}}{m^{2}}\Biggl(\frac{1}{2}\left(ma\right)K_{1}\left(ma\right)-K_{0}\left(ma\right)\Biggr)\nonumber\\
&
&-\frac{\left[\left({\bf{v}}\times{\hat{z}}\right)\cdot{\bf{a}}\right]^{2}}{\left(ma\right)^{2}}\Biggl(\frac{1}{2}\left(ma\right)K_{1}\left(ma\right)+K_{2}\left(ma\right)-\frac{2}{\left(ma\right)^{2}}\Biggr)\nonumber\\
&
&+\frac{{\bf{v}}^{2}}{m^{2}}\Biggl(\frac{1}{2}K_{2}\left(ma\right)-\frac{1}{\left(ma\right)^{2}}\Biggr)\Biggr].
\end{eqnarray}

We notice that the first contribution between brackets in Eq. (\ref{energyEMDD}) is the interaction energy between two Dirac points for the standard Maxwell-Chern-Simons electrodynamics \cite{sourcesMCS}, while the remaining terms account for the Lorentz violation corrections up to order of $\frac{v^{2}}{m^{2}}$. 

As mentioned previously, in the usual Maxwell-Chern-Simons electrodynamics the interaction energy between two Dirac points behaves similarly to that one  between two electric planar charges if we identify, $q_{1}=m\Phi_{1}$ and $q_{2}=m \Phi_{2}$ \cite{sourcesMCS}. Comparing the expressions (\ref{energyEMDD}) and (\ref{energyCCEM2}), we can see that this situation is no longer valid in a Lorentz violation scenario for any nonzero configuration of the background vector.

\begin{figure}[!h]
\centering \includegraphics[scale=0.20]{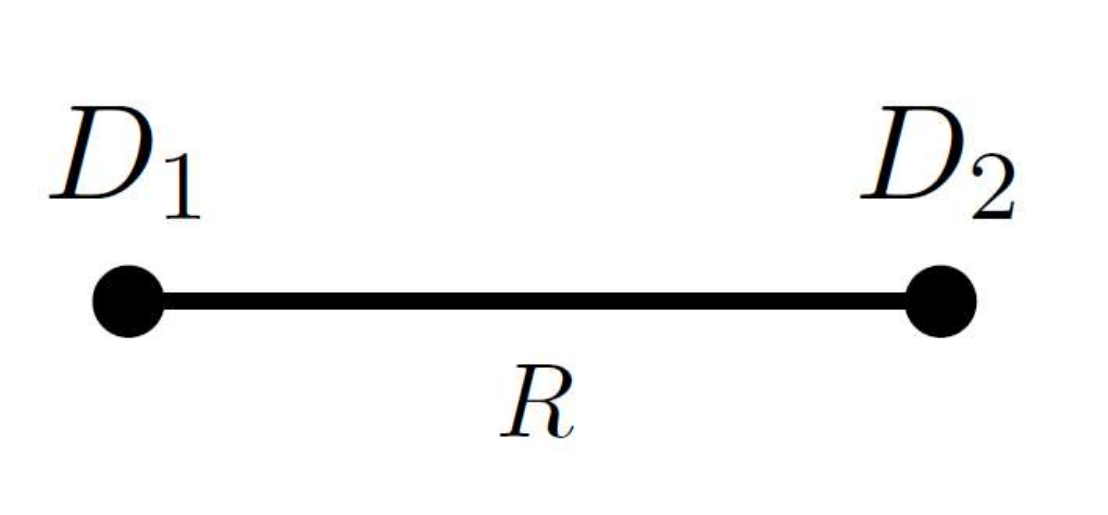} \caption{Two Dirac points separated by a fixed distance $R$. The line is to be interpreted as in Fig.~\ref{figura1}.}
\label{figura9}
\end{figure}

\begin{figure}[!h]
\centering \includegraphics[scale=0.4]{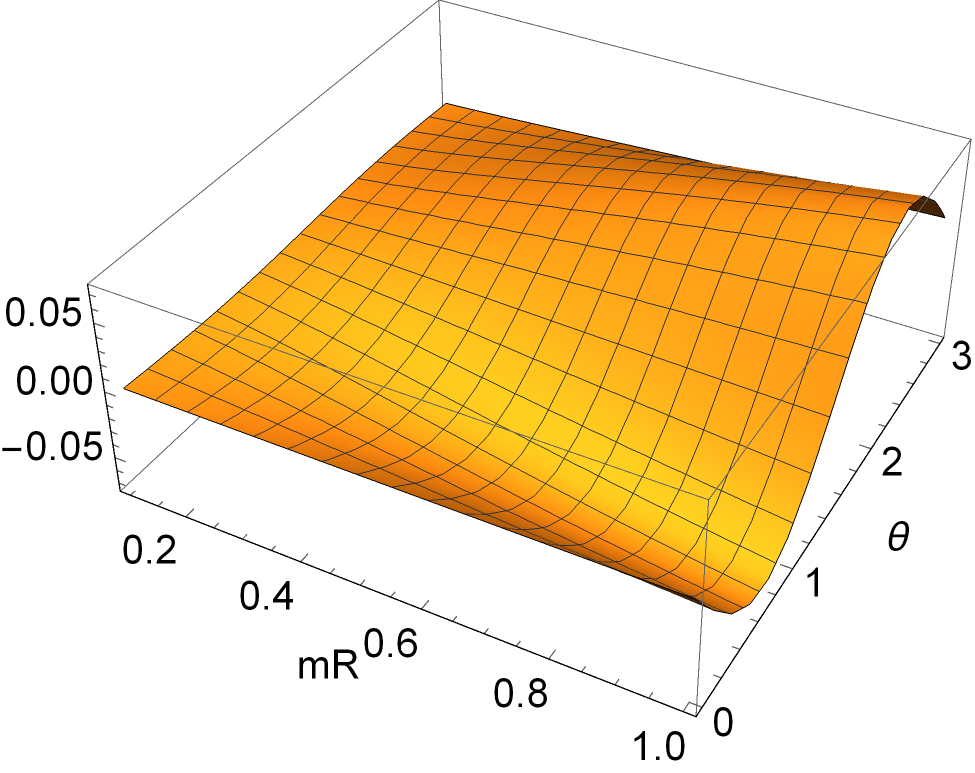} \caption{The torque given in Eq. (\ref{torqueEMDD}), multiplied by $\frac{(2\pi)m^{2}}{m^{2}\Phi_{1}\Phi_{2}{\bf{v}}^{2}}$.}
\label{figura10}
\end{figure}

The interaction energy (\ref{energyEMDD}) provides a Lorentz violating planar physical phenomenon namely, a spontaneous torque on the system shown in Fig. (\ref{figura9}), which is given by
\begin{eqnarray}
\label{torqueEMDD}
\tau_{EM}^{DD}=\frac{m^{2}\Phi_{1}\Phi_{2}}{2\pi}\frac{{\bf{v}}^{2}}{m^{2}}\sin\left(2\theta\right)\Biggl(\frac{1}{2}\left(mR\right)K_{1}\left(mR\right)+K_{2}\left(mR\right)-\frac{2}{\left(mR\right)^{2}}\Biggr) \, .
\end{eqnarray}
The graph in Fig. (\ref{figura10}) shows the general behavior of this torque. 

Finally, for the mixed sector, we have  interactions involving scalar planar charges and Dirac points (let is identify such interactions by the super-index ${\bar{C}}D$), it can be shown that
\begin{eqnarray}
\label{energyM1M2CD}
E_{M1}^{{\bar{C}}D}=E_{M2}^{{\bar{C}}D}&=&-\frac{1}{4\pi}\Biggl[\left(m\sigma_{1}\Phi_{2}+m\sigma_{2}\Phi_{1}\right)\frac{v^{0}}{m}K_{0}\left(ma\right)\nonumber\\
&
&-\left(m\sigma_{1}\Phi_{2}-m\sigma_{2}\Phi_{1}\right)\frac{\left[\left({\bf{v}}\times{\hat{z}}\right)\cdot{\bf{a}}\right]}{ma}\Biggl(K_{1}\left(ma\right)-\frac{1}{ma}\Biggr)\Biggr] \ .
\end{eqnarray}

We can observe that the $(2+1)$-dimensional interaction described by the expression (\ref{energyM1M2CD}) appears due to the presence of the background vector, being of the first order in this parameter. As mentioned previously, in the setup where $v^{0}=0$, the Dirac point behaves similarly to an electric planar charge when it interacts with a scalar planar charge so, if we identify  $q_{1}=m\Phi_{1}$ and $q_{2}=m\Phi_{2}$, Eq. (\ref{energyM1M2CD}) becomes equivalent to (\ref{EM1EM21}). Additionally, there are several particular combinations of parameters for which this interaction disappears, for example, if $v^{0}=0$ and $m\sigma_{1}\Phi_{2}=m\sigma_{2}\Phi_{1}$, also for ${\bf{v}}=0$ and $m\sigma_{1}\Phi_{2}=-m\sigma_{2}\Phi_{1}$, among others.

\begin{figure}[!h]
\centering \includegraphics[scale=0.20]{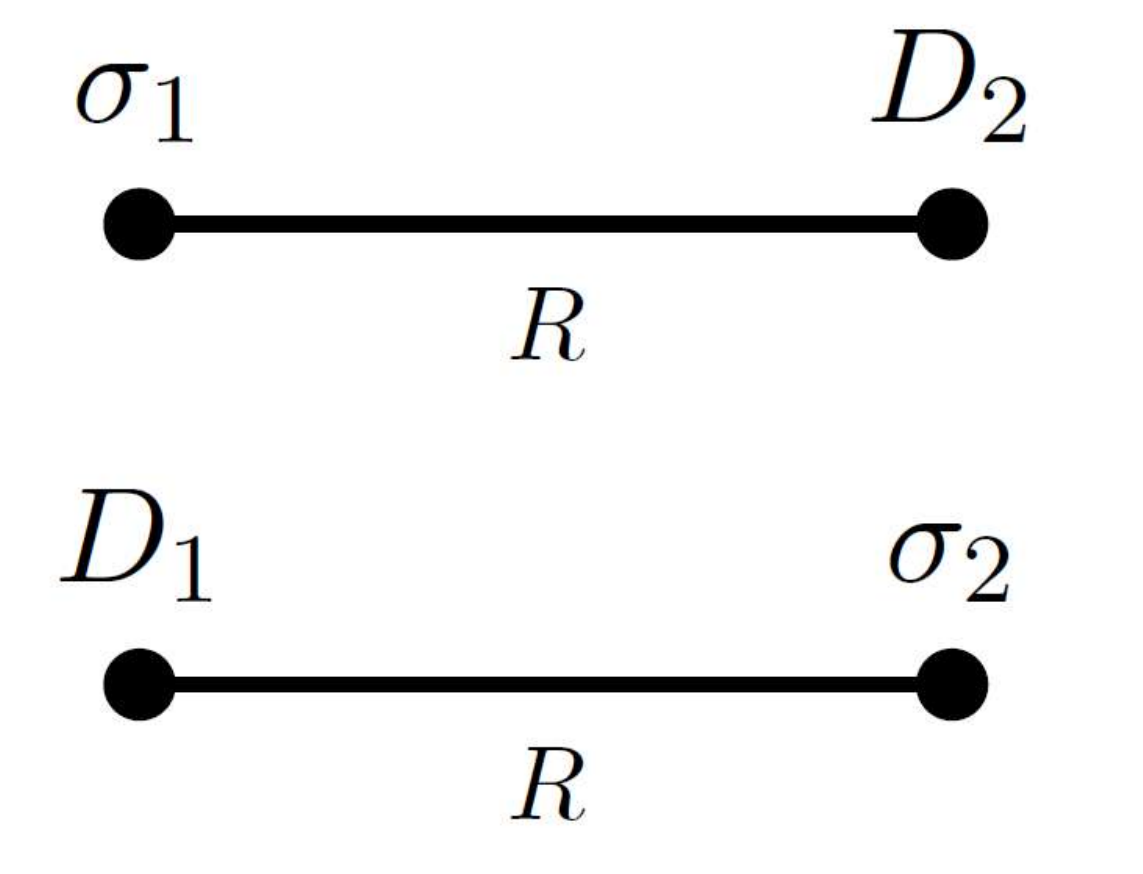} \caption{Representation of the physical system studied in the mixed sector. Here, scalar planar charges and Dirac points are spatially coincident at two points separated by the distance $R$. In the figure, the lines represent the direct interactions between these different types of sources.}
\label{figura11}
\end{figure}

Besides, from Eq. (\ref{energyM1M2CD}) we can obtain a spontaneous torque on the two-dimensional physical system sketched in Fig. (\ref{figura11}),
\begin{eqnarray}
\label{torqueM1M2CD}
\tau_{M}^{{\bar{C}}D}&=&\tau_{M1}^{{\bar{C}}D}+\tau_{M2}^{{\bar{C}}D}\nonumber\\
&=&\frac{\left(m\sigma_{1}\Phi_{2}-m\sigma_{2}\Phi_{1}\right)}{2\pi}\frac{\mid{\bf{v}}\mid}{m}\cos\left(\theta\right)\Biggl(K_{1}\left(mR\right)-\frac{1}{mR}\Biggr) \ ,
\end{eqnarray}
which exhibits a similar behavior to that one shown in Fig. (\ref{figura5}). If $\mid{\bf{v}}\mid=0$ or $m\sigma_{1}\Phi_{2}=m\sigma_{2}\Phi_{1}$ this physical phenomenon vanishes.

\section{ELECTROMAGNETIC FIELD}
\label{EM}

We start this section by computing  the electromagnetic field configuration produced by a Dirac point placed at the origin at an arbitrary point represented by the two-dimensional vector ${\bf{r}}$.  

The electric field can be calculated from the Eq. (\ref{EMA0D}) obtained in the appendix \ref{A}, as follows
\begin{eqnarray}
\label{eleFD}
{\bf{E}}_{(D)}&=&-{\bf{\nabla}}_{\bf{r}}A^{0}_{(D)}\left(\mid{\bf{r}}\mid\right) \nonumber\\
&=&\frac{m^{2}\Phi}{2\pi}K_{1}\left(m\mid{\bf{r}}\mid\right){\hat{r}}+\Delta{\bf{E}}_{(D)} \ ,
\end{eqnarray}
with
\begin{eqnarray}
\label{eleFD33}
\Delta{\bf{E}}_{(D)}&=&-\frac{m^{2}\Phi}{2\pi}\Biggl\{\Biggl[-\frac{v^{0}}{m}\frac{\left[\left({\bf{v}}\times{\hat{z}}\right)\cdot{\bf{r}}\right]}{m\mid{\bf{r}}\mid}\Biggl(\frac{2}{\left(m\mid{\bf{r}}\mid\right)^{2}}-K_{2}\left(m\mid{\bf{r}}\mid\right)\Biggr)\nonumber\\
&
&-\frac{\left(v^{0}\right)^{2}}{2m^{2}}\left(m\mid{\bf{r}}\mid\right)K_{0}\left(m\mid{\bf{r}}\mid\right)+\frac{\left[\left({\bf{v}}\times{\hat{z}}\right)\cdot{\bf{r}}\right]^{2}}{\left(m\mid{\bf{r}}\mid\right)^{2}}\Biggl(\frac{[8+\left(m\mid{\bf{r}}\mid\right)^{2}]}{2\left(m\mid{\bf{r}}\mid\right)}K_{0}\left(m\mid{\bf{r}}\mid\right)\nonumber\\
&
&+\frac{2[4+\left(m\mid{\bf{r}}\mid\right)^{2}]}{\left(m\mid{\bf{r}}\mid\right)^{2}}K_{1}\left(m\mid{\bf{r}}\mid\right)\Biggr)-\frac{{\bf{v}}^{2}}{4m^{2}}\left[K_{1}\left(m\mid{\bf{r}}\mid\right)+K_{3}\left(m\mid{\bf{r}}\mid\right)\right]\Biggr]{\hat{r}}\nonumber\\
&
&-\Biggl[\frac{v^{0}}{m}\Biggl(\frac{K_{1}\left(m\mid{\bf{r}}\mid\right)}{m\mid{\bf{r}}\mid}-\frac{1}{\left(m\mid{\bf{r}}\mid\right)^{2}}\Biggr)+2\frac{\left[\left({\bf{v}}\times{\hat{z}}\right)\cdot{\bf{r}}\right]}{m\mid{\bf{r}}\mid}\nonumber\\
&
&\times\Biggl(\frac{1}{2}K_{1}\left(m\mid{\bf{r}}\mid\right)+\frac{K_{2}\left(m\mid{\bf{r}}\mid\right)}{m\mid{\bf{r}}\mid}\Biggr)\Biggr]\left(\frac{{\bf{v}}}{m}\times{\hat{z}}\right)\Biggr\} \ .
\end{eqnarray}

In the expression (\ref{eleFD}) the first term on the right side is the electric field obtained in standard Maxwell-Chern-Simons electrodynamics \cite{sourcesMCS}, the second one, given by (\ref{eleFD33}), is a  correction that appears due to the presence of the background vector up to order of $\frac{v^{2}}{m^{2}}$. 

\begin{figure}[!h]
\centering \includegraphics[scale=0.4]{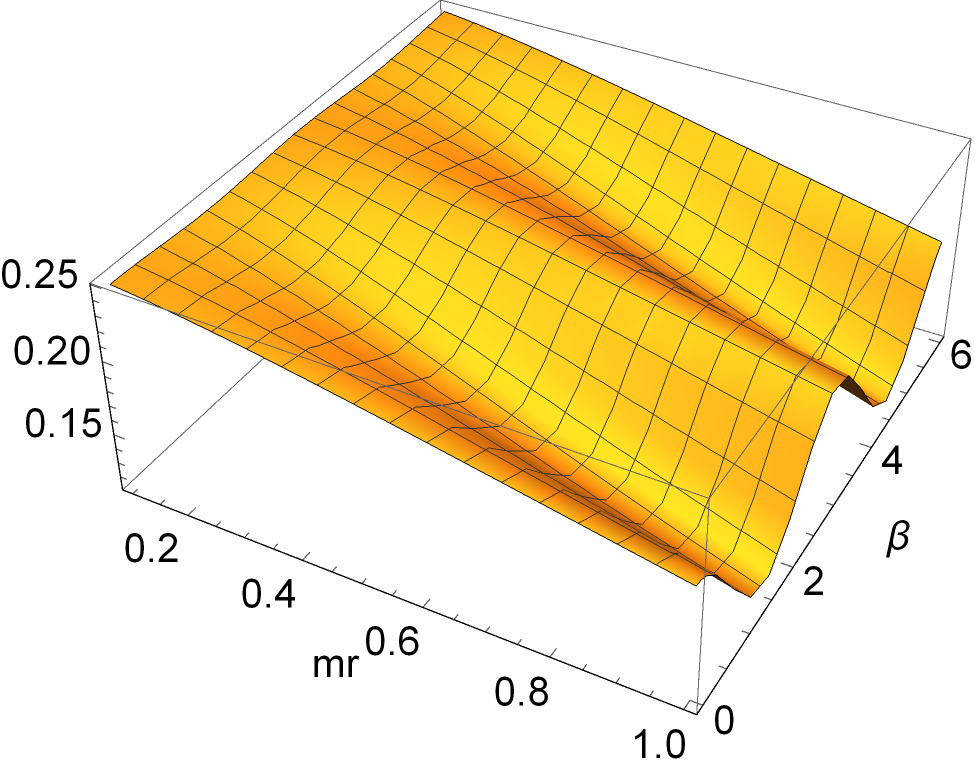} \caption{Plot for $\frac{(2\pi)m^{2}}{m^{2}\Phi{\bf{v}}^{2}}\Delta B_{(D)}$, given in Eq. (\ref{deltaBD}), in the setup where $v^{\mu}=\left(0,{\bf{v}}\right)$.}
\label{figura12}
\end{figure}

From the Eq. (\ref{EMAVD}) (obtained in the appendix \ref{A}), the magnetic field reads
\begin{eqnarray}
\label{BD}
B_{(D)}&=&\epsilon^{0 i j}\frac{\partial A_{(D)}^{j}}{\partial r^{i}}\nonumber\\
&=&-\frac{m^{2}\Phi}{2\pi}K_{0}\left(m\mid{\bf{r}}\mid\right)+\Delta B_{(D)} \ ,
\end{eqnarray}
where we have the magnetic field produced by a Dirac point in the usual Maxwell-Chern-Simons electrodynamics \cite{sourcesMCS} added by a Lorentz violation correction  $\Delta B_{(D)}$ up to order of $\frac{v^{2}}{m^{2}}$, 
\begin{eqnarray}
\label{deltaBD}
\Delta B_{(D)}&=&\frac{m^{2}\Phi}{2\pi}\Biggl[\frac{\left(v^{0}\right)^{2}}{m^{2}}\Biggl(K_{0}\left(m\mid{\bf{r}}\mid\right)-\frac{1}{2}\left(m\mid{\bf{r}}\mid\right)K_{1}\left(m\mid{\bf{r}}\mid\right)\Biggr)\nonumber\\
&
&+\frac{\left[\left({\bf{v}}\times{\hat{z}}\right)\cdot{\bf{r}}\right]^{2}}{\left(m\mid{\bf{r}}\mid\right)^{2}}\Biggl(K_{0}\left(m\mid{\bf{r}}\mid\right)
+\frac{[4+\left(m\mid{\bf{r}}\mid\right)^{2}]}{2\left(m\mid{\bf{r}}\mid\right)}K_{1}\left(m\mid{\bf{r}}\mid\right)-\frac{2}{\left(m\mid{\bf{r}}\mid\right)^{2}}\Biggr)\nonumber\\
&
&+\frac{{\bf{v}}^{2}}{m^{2}}\Biggl(\frac{1}{\left(m\mid{\bf{r}}\mid\right)^{2}}-\frac{1}{2}K_{2}\left(m\mid{\bf{r}}\mid\right)\Biggl)\Biggr] \ .
\end{eqnarray}
Defining by $\beta\in[0,2\pi)$ the angle between ${\bf{v}}$ and ${\bf{r}}$, in Fig. (\ref{figura12}) we have a plot representing the behavior of $\Delta B_{(D)}$ in the setup where $v^{\mu}=\left(0,{\bf{v}}\right)$. 

Now, for a point-like planar charge concentrated at the origin, from the results in the appendix \ref{A}, we have
\begin{eqnarray}
\label{EC}
{\bf{{E}}}_{(C)}&=&\frac{mq}{2\pi}K_{1}\left(m\mid{\bf{r}}\mid\right){\hat{r}}+\Delta{\bf{E}}_{(C)} \ ,
\end{eqnarray}
with
\begin{eqnarray}
\label{ECDELTA}
\Delta{\bf{E}}_{(C)}&=&-\frac{mq}{2\pi}\Biggl\{\Biggl[\frac{\left(v^{0}\right)^{2}}{m^{2}}\Biggl(\frac{1}{m\mid{\bf{r}}\mid}-\frac{1}{2}\left(m\mid{\bf{r}}\mid\right)K_{2}\left(m\mid{\bf{r}}\mid\right)\Biggr)\nonumber\\
&
&+\frac{\left[\left({\bf{v}}\times{\hat{z}}\right)\cdot{\bf{r}}\right]^{2}}{\left(m\mid{\bf{r}}\mid\right)^{2}}\Biggl(\frac{[8+\left(m\mid{\bf{r}}\mid\right)^{2}]}{2\left(m\mid{\bf{r}}\mid\right)}K_{0}\left(m\mid{\bf{r}}\mid\right)+\frac{2[4+\left(m\mid{\bf{r}}\mid\right)^{2}]}{\left(m\mid{\bf{r}}\mid\right)^{2}}K_{1}\left(m\mid{\bf{r}}\mid\right)\Biggr)\nonumber\\
&
&-\frac{{\bf{v}}^{2}}{4m^{2}}\left[K_{1}\left(m\mid{\bf{r}}\mid\right)+K_{3}\left(m\mid{\bf{r}}\mid\right)\right]\Biggr]{\hat{r}}\nonumber\\
&
&-\frac{2\left[\left({\bf{v}}\times{\hat{z}}\right)\cdot{\bf{r}}\right]}{m\mid{\bf{r}}\mid}\Biggl(\frac{1}{2}K_{1}\left(m\mid{\bf{r}}\mid\right)+\frac{K_{2}\left(m\mid{\bf{r}}\mid\right)}{m\mid{\bf{r}}\mid}\Biggr)\Bigl(\frac{{\bf{v}}}{m}\times{\hat{z}}\Bigr)\Biggr\} \ ,
\end{eqnarray}
where we have the electric field generated by an electric planar charge in the standard Maxwell-Chern-Simons electrodynamics \cite{sourcesMCS} plus Lorentz violation contributions up to order of $\frac{v^{2}}{m^{2}}$, which are contained in $\Delta{\bf{E}}_{(C)}$.

Taking the setup where $v^{0}=0$ and identifying $q=m\Phi$, the electric field produced by the Dirac point in Eq. (\ref{eleFD}) becomes equivalent to that one produced by an electric planar charge in (\ref{EC}), showing that for this particular case the Dirac point behaves like a point-planar charge for the electric field configuration.

With the aid of the expression (\ref{ACVF}) the magnetic field is given by
\begin{eqnarray}
\label{BC}
B_{(C)}=-\frac{mq}{2\pi}K_{0}\left(m\mid{\bf{r}}\mid\right)+\Delta B_{(C)} \ ,
\end{eqnarray}
where
\begin{eqnarray}
\label{deltaBC}
\Delta B_{(C)}&=&\frac{mq}{2\pi}\Biggl[-\frac{\left(v^{0}\right)^{2}}{2m^{2}}\left(m\mid{\bf{r}}\mid\right)K_{1}\left(m\mid{\bf{r}}\mid\right)-\frac{v^{0}}{m}\frac{\left[\left({\bf{v}}\times{\hat{z}}\right)\cdot{\bf{r}}\right]}{m\mid{\bf{r}}\mid}\Biggl(K_{1}\left(m\mid{\bf{r}}\mid\right)-\frac{1}{m\mid{\bf{r}}\mid}\Biggr)\nonumber\\
&
&-\frac{{\bf{v}}^{2}}{2m^{2}}K_{2}\left(m\mid{\bf{r}}\mid\right)+\frac{\left[\left({\bf{v}}\times{\hat{z}}\right)\cdot{\bf{r}}\right]^{2}}{\left(m\mid{\bf{r}}\mid\right)^{2}}\Biggl(K_{0}\left(m\mid{\bf{r}}\mid\right)+\frac{[4+\left(m\mid{\bf{r}}\mid\right)^{2}]}{2\left(m\mid{\bf{r}}\mid\right)}K_{1}\left(m\mid{\bf{r}}\mid\right)\Biggr)\Biggr].
\end{eqnarray}

\begin{figure}[!h]
\centering \includegraphics[scale=0.4]{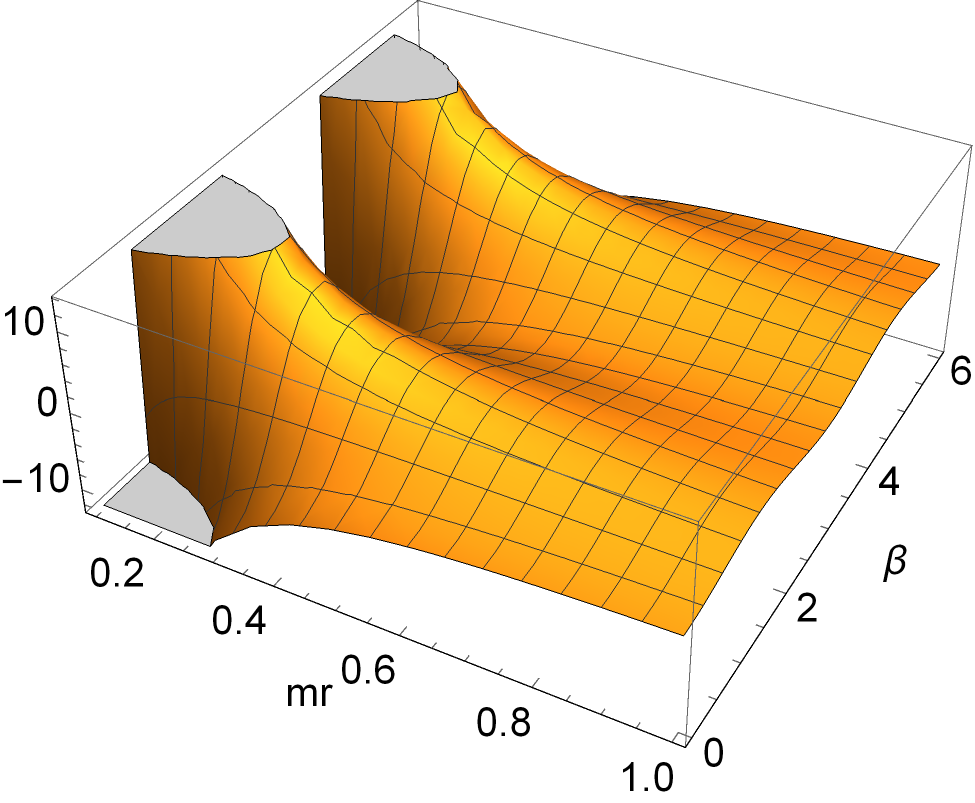} \caption{Plot for $\Delta B_{(C)}$ in Eq. (\ref{deltaBC}), multiplied by $\frac{(2\pi)m^{2}}{mq{\bf{v}}^{2}}$, in the setup where $v^{\mu}=\left(0,{\bf{v}}\right)$.}
\label{figura13}
\end{figure}

We notice that the magnetic field (\ref{BC}) is composed by two parts, the first one stands for the standard result computed in Maxwell-Chern-Simons theory \cite{sourcesMCS}, and the second one, given by (\ref{deltaBC}), contains contributions due to the presence of the background vector up to order of $\frac{v^{2}}{m^{2}}$. In Fig. (\ref{figura13}) we have a plot showing the behavior of the expression (\ref{deltaBC}) for the particular case where $v^{\mu}=\left(0,{\bf{v}}\right)$, which is a similar situation for the one studied in Ref. \cite{clodd}. In this case we highlight that $B_{(C)}\left(v^{0}=0\right)=-mA^{0}_{(C)}\left(\mid{\bf{r}}\mid\right)$.

Unlike what happens in the electric field configuration, comparing the expressions (\ref{BD}) and (\ref{BC}), we notice that the Dirac point behaves as an electric planar charge in the magnetic field configuration only in the absence of the background vector (usual Maxwell-Chern-Simons theory) \cite{sourcesMCS}.

\section{CONCLUSIONS AND PERSPECTIVES}
\label{conclusions}

In this paper we investigated some classical Lorentz violating physical phenomena in $(2+1)$-dimensional electrodynamics due to the presence of stationary point-like field sources. 

We started from the Carroll-Field-Jackiw model defined in $(3+1)$ dimensions, belonging to the gauge CPT-odd sector of the SME, and apply the dimensional reduction procedure, resulting in a planar model composed by an electromagnetic sector represented by the Maxwell-Chern-Simons electrodynamics, a Klein-Gordon sector  and a mixed one. For all the sectors of this theory, we explored some physical effects, which emerged from the interactions between external sources.

We have computed perturbative results up to second order in the background vector related to the presence of both electric and scalar planar charges and Dirac points. 
Specifically, we obtained physical phenomena such as spontaneous torques, interaction forces and electromagnetic fields.

It would be interesting to carry out a similar investigation for the $(2+1)$-dimensional CPT-odd gauge sector of the nonminimal SME  \cite{pnmSME,work}.

{\bf Acknowledgements.} 
This study was financed in part by the Coordena\c c\~ao de Aperfei\c coamento de Pessoal de N\'\i vel Superior -- Brasil (CAPES) -- Finance Code 001 (LHCB and PHOS). AFF acknowledges the support of Conselho Nacional de Desenvolvimento Cient\'\i fico e Tecnol\'ogico (CNPq) via the grant 305967/2020-7. FAB thanks to CNPq  under the grants 313426/2021-0 for financial support

\appendix                                     
\section{FIELD CONFIGURATIONS}
\label{A}
In this appendix we compute the field configuration  generated by both the Dirac point and the electric planar charge.

The field solution produced by a given external source can be obtained from the expression,
\begin{eqnarray}
\label{gaugeF}
A^{\mu}\left(r\right)&=&\int d^{3}y D^{\mu\nu}\left(r,y\right)J_{\nu}\left(y\right)\nonumber\\
&=&\int\frac{d^{3}p}{\left(2\pi\right)^{3}}\int d^{3}y \ e^{-ip\cdot\left(r-y\right)}{\tilde{D}}^{\mu\nu}\left(p\right)J_{\nu}\left(y\right)  \ .
\end{eqnarray}

For a Dirac point placed at origin, we have
\begin{eqnarray}
\label{sourceD0}
J_{\nu(D)}\left({\bf y}\right)=-2\pi i\Phi\int\frac{d^{3}p}{\left(2\pi\right)^{3}} \ 
\delta\left(p^{0}\right)\epsilon_{0\nu\sigma}p^{\sigma} \ 
e^{-i p\cdot y} \ .
\end{eqnarray}
Substituting (\ref{propem}) and (\ref{sourceD0}) in Eq.  (\ref{gaugeF}), and then performing the calculations, we arrive at
\begin{eqnarray}
\label{EMA0D}
A^{0}_{(D)}\left(\mid{\bf{r}}\mid\right)&=&\frac{m\Phi}{2\pi}\Biggl[K_{0}\left(m\mid{\bf{r}}\mid\right)-\frac{v^{0}}{m}\frac{\left[\left({\bf{v}}\times{\hat{z}}\right)\cdot{\bf{r}}\right]}{m\mid{\bf{r}}\mid}\Biggl(K_{1}\left(m\mid{\bf{r}}\mid\right)-\frac{1}{m\mid{\bf{r}}\mid}\Biggr)\nonumber\\
&
&+\frac{\left(v^{0}\right)^{2}}{2m^{2}}\left(m\mid{\bf{r}}\mid\right)K_{1}\left(m\mid{\bf{r}}\mid\right)-\frac{\left[\left({\bf{v}}\times{\hat{z}}\right)\cdot{\bf{r}}\right]^{2}}{\left(m\mid{\bf{r}}\mid\right)^{2}}\nonumber\\
&
&\times\left(\frac{1}{2}\left(m\mid{\bf{r}}\mid\right)K_{1}\left(m\mid{\bf{r}}\mid\right)+K_{2}\left(m\mid{\bf{r}}\mid\right)\right)
+\frac{{\bf{v}}^{2}}{2m^{2}}K_{2}\left(m\mid{\bf{r}}\mid\right)\Biggr] \ ,
\end{eqnarray}
and
\begin{eqnarray}
\label{EMAVD}
{\bf{A}}_{(D)}\left({\bf{r}}\right)&=&\frac{m\Phi}{2\pi}\Biggl\{\Biggl[-K_{1}\left(m\mid{\bf{r}}\mid\right)-\frac{\left(v^{0}\right)^{2}}{2m^{2}}\left(m\mid{\bf{r}}\mid\right)K_{0}\left(m\mid{\bf{r}}\mid\right)+\frac{v^{0}}{m}\frac{\left[\left({\bf{v}}\times{\hat{z}}\right)\cdot{\bf{r}}\right]}{m\mid{\bf{r}}\mid}\nonumber\\
&
&\times\Biggl(K_{2}\left(m\mid{\bf{r}}\mid\right)+\frac{1}{2}\left(m\mid{\bf{r}}\mid\right)K_{1}\left(m\mid{\bf{r}}\mid\right)\Biggr)\Biggr]\left({\hat{r}}\times{\hat{z}}\right)\nonumber\\
&
&+\Biggl[\frac{v^{0}}{m}\Biggl(K_{2}\left(m\mid{\bf{r}}\mid\right)+\frac{1}{2}\left(m\mid{\bf{r}}\mid\right)K_{1}\left(m\mid{\bf{r}}\mid\right)\Biggr)
-\frac{\left[\left({\bf{v}}\times{\hat{z}}\right)\cdot{\bf{r}}\right]}{m\mid{\bf{r}}\mid}\nonumber\\
&
&\times\Biggl(-\frac{1}{2\left(m\mid{\bf{r}}\mid\right)}+\frac{3}{2}K_{1}\left(m\mid{\bf{r}}\mid\right)+\frac{1}{2}\left(m\mid{\bf{r}}\mid\right)K_{0}\left(m\mid{\bf{r}}\mid\right)+2\frac{K_{2}\left(m\mid{\bf{r}}\mid\right)}{m\mid{\bf{r}}\mid}\Biggr)\Biggr]\frac{{\bf{v}}}{m}\nonumber\\
&
&-\frac{\left({\bf{v}}\cdot{\bf{r}}\right)}{m\mid{\bf{r}}\mid}\Biggl(\frac{1}{2\left(m\mid{\bf{r}}\mid\right)}+\frac{1}{2}K_{1}\left(m\mid{\bf{r}}\mid\right)+2\frac{K_{2}\left(m\mid{\bf{r}}\mid\right)}{m\mid{\bf{r}}\mid}\Biggr)\left(\frac{{\bf{v}}}{m}\times{\hat{z}}\right)\nonumber\\
&
&-\Biggl[\frac{v^{0}}{m}\frac{\left({\bf{v}}\cdot{\bf{r}}\right)}{m\mid{\bf{r}}\mid}
\Biggl(K_{2}\left(m\mid{\bf{r}}\mid\right)+\frac{1}{2}\left(m\mid{\bf{r}}\mid\right)K_{1}\left(m\mid{\bf{r}}\mid\right)\Biggr)
-\frac{\left({\bf{v}}\cdot{\bf{r}}\right)}{m\mid{\bf{r}}\mid}\frac{\left[\left({\bf{v}}\times{\hat{z}}\right)\cdot{\bf{r}}\right]}{m\mid{\bf{r}}\mid}\nonumber\\
&
&\times\Biggl(\frac{1}{m\mid{\bf{r}}\mid}+\frac{1}{2}\left(m\mid{\bf{r}}\mid\right)K_{0}\left(m\mid{\bf{r}}\mid\right)+3K_{1}\left(m\mid{\bf{r}}\mid\right)+8\frac{K_{2}\left(m\mid{\bf{r}}\mid\right)}{m\mid{\bf{r}}\mid}\Biggr)\Biggr]{\hat{r}}\Biggr\} \ ,
\end{eqnarray}
where ${\hat{r}}$ is the unit vector pointing on the direction of ${\bf{r}}$.

We notice that the above potentials are static and were computed perturbatively up to order of $\frac{v^{2}}{m^{2}}$, they give an electric field as well as a magnetic field produced by a Dirac point. In the absence of the background vector, we recover the potentials obtained in the usual Maxwell-Chern-Simons electrodynamics \cite{sourcesMCS}.

Now, for an electric planar charge placed at origin namely, 
$J_{\nu(C)}\left(y\right)=q\eta_{\nu}^{\ 0}\delta^{2}\left({\bf{y}}\right)$, we obtain that
\begin{eqnarray}
\label{A0C}
A^{0}_{(C)}\left(\mid{\bf{r}}\mid\right)&=&\frac{q}{2\pi}\Biggl[K_{0}\left(m\mid{\bf{r}}\mid\right)+\frac{\left(v^{0}\right)^{2}}{m^{2}}\left(\ln\left(\frac{\mid{\bf{r}}\mid}{r_{0}}\right)+K_{0}\left(m\mid{\bf{r}}\mid\right)+\frac{1}{2}\left(m\mid{\bf{r}}\mid\right)K_{1}\left(m\mid{\bf{r}}\mid\right)\right)\nonumber\\
&
&-\frac{\left[\left({\bf{v}}\times{\hat{z}}\right)\cdot{\bf{r}}\right]^{2}}{\left(m\mid{\bf{r}}\mid\right)^{2}}\left(\frac{1}{2}\left(m\mid{\bf{r}}\mid\right)K_{1}\left(m\mid{\bf{r}}\mid\right)+K_{2}\left(m\mid{\bf{r}}\mid\right)\right)+\frac{{\bf{v}}^{2}}{2m^{2}}K_{2}\left(m\mid{\bf{r}}\mid\right)\Biggr] \ ,
\end{eqnarray}
and
\begin{eqnarray}
\label{ACVF}
{\bf{A}}_{(C)}\left({\bf{r}}\right)&=&\frac{q}{2\pi}\Biggl\{\Biggl[\frac{1}{m\mid{\bf{r}}\mid}-K_{1}\left(m\mid{\bf{r}}\mid\right)-\frac{\left(v^{0}\right)^{2}}{2m^{2}}\left(m\mid{\bf{r}}\mid\right)K_{2}\left(m\mid{\bf{r}}\mid\right)+\frac{v^{0}}{m}\frac{\left[\left({\bf{v}}\times{\hat{z}}\right)\cdot{\bf{r}}\right]}{m\mid{\bf{r}}\mid}\nonumber\\
&
&\times\Biggl(K_{2}\left(m\mid{\bf{r}}\mid\right)+\frac{1}{2}\left(m\mid{\bf{r}}\mid\right)K_{1}\left(m\mid{\bf{r}}\mid\right)\Biggr)\Biggr]\left({\hat{r}}\times{\hat{z}}\right)+\Biggl[\frac{v^{0}}{2m}\Biggl(\ln\left(\frac{\mid{\bf{r}}\mid}{r_{0}}\right)\nonumber\\
&
&+K_{0}\left(m\mid{\bf{r}}\mid\right)+\left(m\mid{\bf{r}}\mid\right)K_{1}\left(m\mid{\bf{r}}\mid\right)+3K_{2}\left(m\mid{\bf{r}}\mid\right)\Biggr)
-\frac{\left[\left({\bf{v}}\times{\hat{z}}\right)\cdot{\bf{r}}\right]}{2\left(m\mid{\bf{r}}\mid\right)}\nonumber\\
&
&\times\Biggl(K_{1}\left(m\mid{\bf{r}}\mid\right)+\left(m\mid{\bf{r}}\mid\right)K_{2}\left(m\mid{\bf{r}}\mid\right)+4\frac{K_{2}\left(m\mid{\bf{r}}\mid\right)}{m\mid{\bf{r}}\mid}\Biggr)\Biggr]\frac{{\bf{v}}}{m}\nonumber\\
&
&-\frac{\left({\bf{v}}\cdot{\bf{r}}\right)}{m\mid{\bf{r}}\mid}\Biggl(\frac{1}{2}K_{1}\left(m\mid{\bf{r}}\mid\right)+2\frac{K_{2}\left(m\mid{\bf{r}}\mid\right)}{m\mid{\bf{r}}\mid}\Biggr)\left(\frac{{\bf{v}}}{m}\times{\hat{z}}\right)
-\Biggl[\frac{v^{0}}{m}\frac{\left({\bf{v}}\cdot{\bf{r}}\right)}{m\mid{\bf{r}}\mid}\nonumber\\
&
&\times\Biggl(\frac{1}{2}+\frac{1}{2}\left(m\mid{\bf{r}}\mid\right)K_{1}\left(m\mid{\bf{r}}\mid\right)+2K_{2}\left(m\mid{\bf{r}}\mid\right)\Biggr)
-\frac{\left({\bf{v}}\cdot{\bf{r}}\right)}{m\mid{\bf{r}}\mid}\frac{\left[\left({\bf{v}}\times{\hat{z}}\right)\cdot{\bf{r}}\right]}{m\mid{\bf{r}}\mid}\nonumber\\
&
&\times\Biggl(\frac{1}{2}\left(m\mid{\bf{r}}\mid\right)K_{0}\left(m\mid{\bf{r}}\mid\right)+3K_{1}\left(m\mid{\bf{r}}\mid\right)+8\frac{K_{2}\left(m\mid{\bf{r}}\mid\right)}{m\mid{\bf{r}}\mid}\Biggr)\Biggr]{\hat{r}}\Biggr\} \ ,
\end{eqnarray}

On the right hand side of both expressions (\ref{A0C}) and (\ref{ACVF}) the independent contributions of the background vector give the well-known results in the literature for the usual Maxwell-Chern-Simons electrodynamics \cite{sourcesMCS}, the dependent ones are corrections up to order of $\frac{v^{2}}{m^{2}}$.



\end{document}